\documentclass[sigconf]{acmart}
\usepackage{subcaption}
\newcommand{\omvs}{OMVs\xspace}
\newcommand{\omv}{OMV\xspace}
\newcommand{\gog}{GoG\xspace}
\newcommand{\otherattr}{attribute-2\xspace} % double check
\newcommand{\cref}[1]{\autoref{#1}}

\newcommand{\EplusM}{\texttt{EplusM}\xspace}
\newcommand{\vcfont}[1]{\textsf{#1}}
\newcommand{\eg}{e.g.,\xspace}
\newcommand{\ie}{i.e.,\xspace}
\newcommand{\posx}{\vcfont{PosX}\xspace}
\newcommand{\posy}{\vcfont{PosY}\xspace}
\newcommand{\row}{\vcfont{Row}\xspace}
\newcommand{\col}{\vcfont{Col}\xspace}
\newcommand{\len}{\vcfont{Length}\xspace}
\newcommand{\area}{\vcfont{Area}\xspace}
\newcommand{\colint}{\vcfont{Intensity}\xspace}
\newcommand{\colhue}{\vcfont{Hue}\xspace}
\newcommand{\shape}{\vcfont{Shape}\xspace}
\newcommand{\VC}{\textrm{VC}\xspace}
\newcommand{\LRE}{$e_{\log}$\xspace}

\newcommand{\SSB}{\protect\hyperlink{SSB}{SSB}\xspace}
\newcommand{\WSB}{\protect\hyperlink{WSB}{WSB}\xspace}
\newcommand{\OMC}{\protect\hyperlink{OMC}{OMC}\xspace}
\newcommand{\OML}{\protect\hyperlink{OML}{OML}\xspace}
\newcommand{\OMH}{\protect\hyperlink{OMH}{OMH}\xspace}
\newcommand{\OMM}{\protect\hyperlink{OMM}{OMM}\xspace}

\graphicspath{{img/}} % where to search for the images

\AtBeginDocument{%
  }

\copyrightyear{2025}
\acmYear{2025}
\setcopyright{cc}
\setcctype{by-nc}
\acmConference[CHI '25]{CHI Conference on Human Factors in Computing
Systems}{April 26-May 1, 2025}{Yokohama, Japan}
\acmBooktitle{CHI Conference on Human Factors in Computing Systems (CHI
'25), April 26-May 1, 2025, Yokohama,
Japan}\acmDOI{10.1145/3706598.3713487}
\acmISBN{979-8-4007-1394-1/25/04}

\usepackage{xspace}
\usepackage{eurosym}
\usepackage{comment}
\usepackage{amsmath}
\usepackage{amsfonts}
\usepackage{subcaption}
\usepackage{wrapfig}
\usepackage{graphicx} % Required for including images
\usepackage{makecell} % for new lines in cells and thread
\usepackage{enumitem}
\usepackage{pifont}% http://ctan.org/pkg/pifont
\newcommand{\cmark}{\ding{51}}%
\newcommand{\xmark}{\ding{55}}%
\xspaceaddexceptions{'\}}
\usepackage{xcolor}
\usepackage{pbalance} % balance the two columns on the last page

\definecolor{skyblue}{rgb}{0.467, 0.635, 1}
\definecolor{dustyteal}{rgb}{0.369, 0.596, 0.592}

\newcommand{\added}[1]{\textcolor{black}{#1}}
\newcommand{\changed}[1]{\textcolor{black}{#1}}

\newcommand{\secondChanged}[1]{\textcolor{black}{#1}}

\begin{document}

\title[Lost in Magnitudes: Exploring Visualization Designs for Large Value Ranges]{Lost in Magnitudes: Exploring Visualization Designs\texorpdfstring{\\}{} for Large Value Ranges}

\author{Katerina Batziakoudi}
\orcid{0009-0003-7069-9339}
\affiliation{%
  \institution{Berger-Levrault \& Université Paris-Saclay, CNRS, Inria}
  \city{Orsay}
  \country{France}}
\email{a.batziakoudi@berger-levrault.com}

\author{Florent Cabric}
\orcid{0000-0002-9326-9441}
\affiliation{%
  \institution{Université Paris-Saclay, CNRS, Inria}
  \city{Orsay}
  \country{France}}
\email{florent.cabric@inria.fr}

\author{Stéphanie Rey}
\orcid{0000-0002-2826-2489}
\affiliation{%
  \institution{Berger-Levrault}
  \city{Toulouse}
  \country{France}}
\email{Stephanie.REY@berger-levrault.com}

\author{Jean-Daniel Fekete}
\orcid{0000-0003-3770-8726}
\affiliation{%
  \institution{Université Paris-Saclay, CNRS, Inria}
  \city{Orsay}
  \country{France}}
\email{Jean-Daniel.Fekete@inria.fr}

\renewcommand{\shortauthors}{Batziakoudi et al.}

\begin{abstract}
  We explore the design of visualizations for values spanning multiple orders of magnitude; we call them Orders of Magnitude Values (\omvs).
Visualization researchers have shown that separating \omvs into two components, the \emph{mantissa} and the \emph{exponent}, and encoding them separately overcomes limitations of linear and logarithmic scales. However, only a small number of such visualizations have been tested, and the design guidelines for visualizing the mantissa and exponent separately remain under-explored. 
To initiate this exploration, better understand the factors influencing the effectiveness of these visualizations, and create guidelines, we adopt a multi-stage workflow.
We introduce a design space for visualizing mantissa and exponent, systematically generating and qualitatively evaluating all possible visualizations within it.
From this evaluation, we derive guidelines.
We select two visualizations that align with our guidelines and test them using a crowdsourcing experiment, showing they facilitate quantitative comparisons and increase confidence in interpretation compared to the state-of-the-art.
\end{abstract}




\begin{teaserfigure}
    \centering
    \includegraphics[width=0.85\textwidth]{design-space}
    \caption{ Our design space has three dimensions: MARKS (green), DATA (blue), and VISUAL CHANNELS (red). The figure shows an example of using our design space as an interactive table, where a mark is selected, and visual channels are assigned to data attributes. Grey cells are invalid, according to the visualization literature. After checking for integrity constraints, a visualization is generated. Note the $\bigoplus$ signs (top left); enable the assignment of the \EplusM scale to positions.}
    \label{fig:design-space}
    \Description{The image shows our design space and the defined dimensions. Our design space encompasses three dimensions: MARKS (green), DATA (blue), and VISUAL CHANNELS (red). The image illustrates an example of using our design space as an interactive table, where a mark is selected, and visual channels are assigned to data attributes. Grey cells are invalid, according to the visualization literature. After checking for integrity constraints, a visualization is generated to perform the tasks.}
\end{teaserfigure}


\maketitle

\section{Introduction}

\begin{figure}
    \includegraphics[width=0.49\columnwidth]{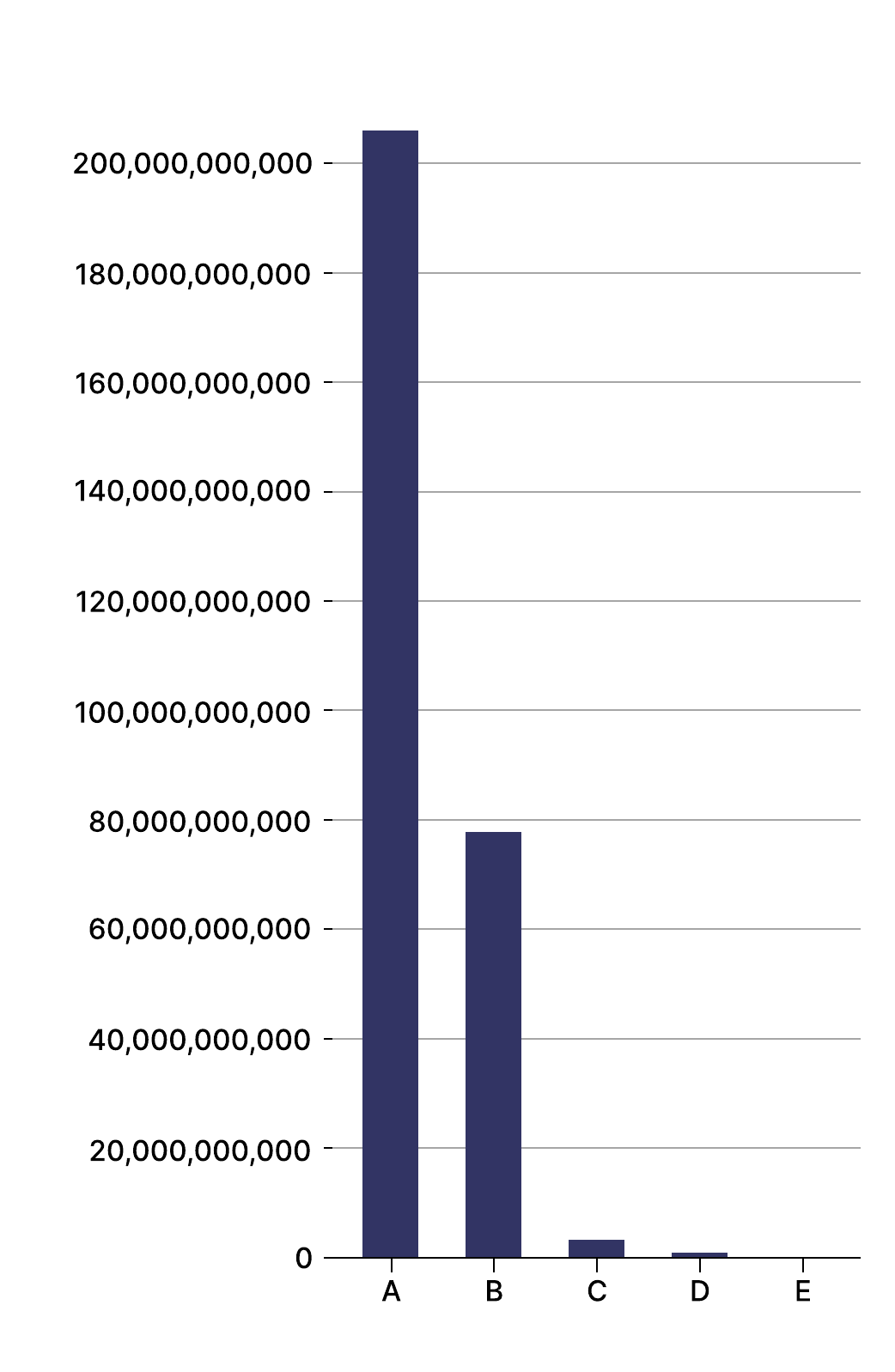}
    \includegraphics[width=0.49\columnwidth]{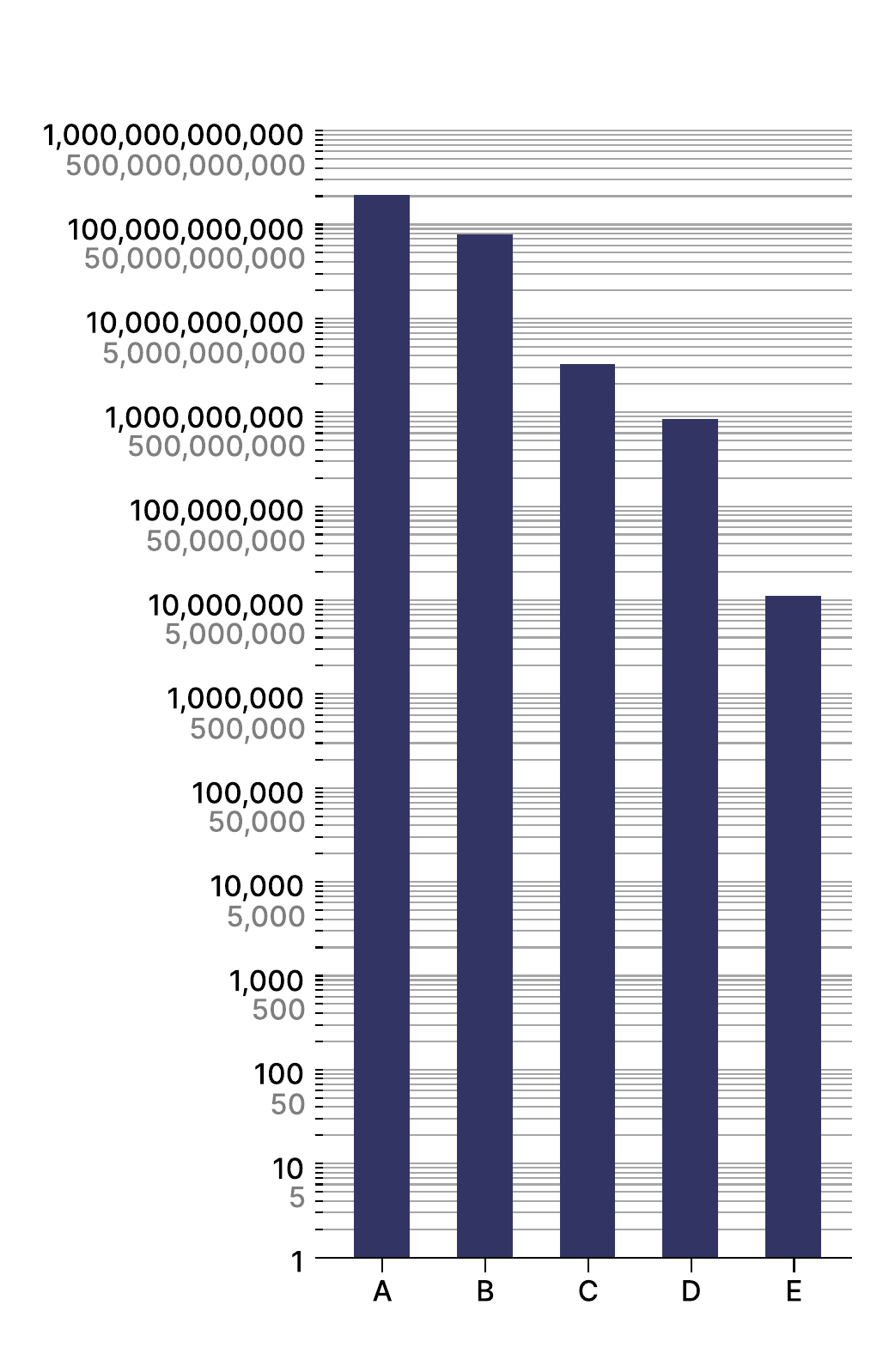}
    \caption{Linear bar chart (left) and logarithmic bar chart (right) illustrating a sample of the French government's budget allocations, showcasing order-of-magnitude differences.}
     \Description{The image shows two bar chart visualizations of selected categories of the budget allocations of the French government, which range from tens of millions of Euros to hundreds of billions of Euros, covering five orders of magnitude. On the left, a linear scale is used, illustrating how lower magnitudes become almost invisible due to the vast differences in value. On the right, a logarithmic scale is used to better represent the full range of values.}
    \label{fig:budget}
    \vspace{-1.5em}
  \end{figure}

In this article, we explore the design of static visualizations with values spanning multiple orders of magnitude, drawing inspiration from scientific notation and floating-point arithmetic that separate values into an exponent and a mantissa part, such that a value $v$ is expressed as $v = \textit{mantissa} \times 10^{\textit{exponent}}$. Order of Magnitude Values (\omvs) are integral to various domains of daily life, including but not limited to financial analysis, pandemic tracking, demographic studies, environmental monitoring, and social media metrics. An illustrative example is the budget allocations of the French government (\cref{fig:budget}), which range from tens of millions of Euros ($10^7$\euro) to hundreds of billions of Euros ($10^{11}$\euro), thereby covering five orders of magnitude. The effective static visualization of \omvs is essential for providing an informative overview suitable for both printed and digital media. 
% We argue that OMVs should be treated as a specific kind of  quantitative data type in visualization, and thus, visualization systems should support these values effectively.

Common visualization techniques, such as linear and logarithmic bar charts, prove ineffective in visualizing \omvs. In the example of the French budget (\cref{fig:budget}), using a linear scale (left) makes it impossible to read and compare budget categories with smaller magnitudes, which are shown with less than one pixel, resulting in information being entirely lost.
While a logarithmic scale (right) can partially mitigate this issue by ensuring that all budget values are visible, it is not well understood by the general public~\cite{ciccione_analyzing_2022, romano_scale_2020, Menge2018}
and may even lead to misconceptions among scientists~\cite{kubina_slope_2023}. Beyond legibility issues, logarithmic scales also hinder efficient quantitative comparisons~\cite{hlawatsch_scale-stack_2013}, such as the low-level tasks of estimating ratios between values within the same or neighboring magnitudes.

 % In scientific notation, the value 5,800,000 is represented as \texttt{5.8E6} where 5.8 is the \emph{mantissa} and 6 the \emph{exponent}. By decomposing large values into these two components---the mantissa and the exponent---we aim to effectively visualize such data within a single view.

Visualization researchers, also inspired by scientific notation, have proposed novel static designs that address the limitations of linear and logarithmic scales~\cite{borgo_order_2014, hohn_width-scale_2020, hlawatsch_scale-stack_2013, braun_color_2022, braun_reclaiming_2023} (see \autoref{fig:omv-visualizations}). They separate \omvs into two parts: the \textit{mantissa} and the \textit{exponent} and visualize each component separately using different visual channels~\cite{hohn_width-scale_2020, hlawatsch_scale-stack_2013, braun_color_2022, braun_reclaiming_2023}, distinct marks~\cite{borgo_order_2014}, or novel scales~\cite{hlawatsch_scale-stack_2013, braun_color_2022, braun_reclaiming_2023}. This decomposition into two parts challenges established graphical perception heuristics for magnitude estimation, as a single quantitative attribute is represented by two interdependent components. Low-level tasks, such as value retrieval and comparison, require the original \omv to be reconstructed by visually combining the mantissa and exponent, introducing novel constraints and requirements. While researchers have empirically demonstrated that this separation is beneficial for visualizing \omvs, only a few visualizations have been explored so far, and the design principles for effectively visualizing mantissa and exponent remain unclear. 

Building on this foundation, to better understand the factors affecting the effectiveness of visualizations that separate \omvs to mantissa and exponent we follow a multistage workflow. We begin by \textbf{reviewing existing literature} on large number estimation, graphical perception, and visualization of large value ranges to \textbf{define design requirements and constraints}. Next, we \textbf{describe a design space}, \added{rooted in the principles of the Grammar of Graphics (\gog)~\cite{GrammarOfGraphics} and} focused on tabular data, the Cartesian system, and visualizations with no redundant encoding. \changed{We \textbf{refine the design space} by applying constraints derived from the literature.}
We \textbf{systematically generate} all possible visualizations within this defined space and we \textbf{qualitatively evaluate} them, aiming to identify encoding and decoding problems that could impact their effectiveness. Building on the results of the qualitative evaluation, we \textbf{derive design guidelines} and \textbf{test them through a crowdsoursing experiment} comparing two newly designed visualizations, aligned with these guidelines, against traditional linear and logarithmic bar charts, as well as a visualization from the state-of-the-art (the Scale-Stacked Bar Chart). Our results demonstrate that the proposed designs performed similarly to, or better than, existing solutions and confirmed that visualizations that both respect our guidelines and separate \omvs into mantissa and exponent facilitate quantitative comparisons.

In summary, we make the following contributions:
\begin{itemize}[nosep]
\item we present a design space for the static visualization of \omvs, when mantissa and exponent are separately visualized
\item we introduce guidelines for designing effective mantissa-exponent visualizations for \omvs
\item \changed{we introduce and validate two visualization designs for \omvs, Facet and \EplusM, that are effective for quantitative tasks.}
\end{itemize}

\changed{Additionally, we address error measurements in \omv experiments to improve the comparability of results in future studies. We also open-sourced the tool developed for our design space exploration, along with all the generated visualizations, to encourage further evaluation and expansion of the defined design space, as well as to support the application of this methodology to other design space exploration challenges. The numerical separation of \omvs into mantissa and exponent has long been valued for its advantages in computational efficiency, numerical stability, and readability across scientific disciplines. Inspired by these strengths, our work demonstrates how this separation can benefit the visualization domain, paving the way for broader adoption and generalization of mantissa-exponent visualizations for \omvs.}

\section{Background}
In this section, we first discuss the definition and numerical relationship between the mantissa and exponent. Then, we present related work in data visualization with varying orders of magnitude and visualization effectiveness.

\subsection{Definition and Terminology}\label{sec:definition}% sounds boring 
We borrow our definitions and terminology from both \href{https://en.wikipedia.org/wiki/Scientific_notation}{\emph{scientific notation}}---a notation for symbolically writing \omvs---and floating-point numbers (see \cite[section 4.2]{SemiNumericalAlgorithms} for a formal introduction to floating point numbers)---a computer representation of \omvs used to perform calculations.
A numerical value $v$ is represented as a pair $(m, e)$ with $v=m \times b^e$, where $m \in \mathbb{R}\setminus\{0\}$ is called \emph{mantissa} (sometimes \emph{significand} or \emph{fraction}) and $e \in \mathbb{Z}$ is called \emph{exponent}. Like scientific notation, we use \emph{base} $b=10$, whereas floating-point numbers use base 2.
% In the following, we refer to the mantissa of $v$ with $M(v) = m$ and the exponent with $E(v) = e$.

In \emph{scientific normalized notation}, $e$ is chosen so that $|m| \in [1,10)$. The mantissa defines the \emph{significant digits} of a value inside an order of magnitude and is represented with a given \emph{precision}, the number of digits $p \in \mathbb{N}^{+}$, that may depend on $v$'s intrinsic precision and on the desired/available display precision. 
% Using the scientific normalized notation, we also call $e$ the \emph{order of magnitude} of a number. 

% or its \emph{level of magnitude}. 

In accordance with Höhn et al.~\cite{hohn_width-scale_2020}, \textbf{a dataset contains orders of magnitude values (\omvs) (or a \emph{large value range}) when one of its numerical attributes contains values that span four orders of magnitude or more.} Using scientific notation and floating point numbers allows for dealing with large ranges of values but also introduces new issues compared to decimal notation~\cite{FloatingPointIssues}. In particular, regarding visualization of values spanning multiple orders of magnitude (hereafter referred to as \emph{\omv visualization}), the order of magnitude of 0 is undefined, and a smaller number can be \emph{negligible} compared to a larger one when their precision is lower than the difference between their exponents. %, \ie $p < E(b)-E(s)$; in that case, $b-s = b$ and $b+s = b$. 
% For example, if $b = 1000$, $s = 1$ and $p=2$, $b+s = 1001$ represented as $1.00e3$, just like $b$, hence $s$ is negligible compared to $b$. 
% Too early
% When visualized, the same issue arises due to the limited accuracy of the visual channels, limiting the precision of value lookup and comparison tasks.
% Formally, it is when $E(B)-E(S) > p+2$ due to rounding rules

 \subsection{Visualization and Perception of Large Value Ranges} 
    In the context of data visualization, visually representing \omvs poses significant challenges. While linear scales are widely understood, they fall short of effectively depicting large value ranges, rendering only the larger values legibly. If the screen has a width and height in the $10^3$ pixels,
    linear scale is not capable of visualizing values that are more than three orders of magnitude apart, as shown in \autoref{fig:budget}. This distinctiveness issue~\cite{rensink2013prospects} hinders fundamental analytical tasks, including value estimation and comparison (ratio and difference) between different ranges. Although dual scale charts~\cite{Dual-Scale} and innovative designs like the Du Bois Wrapped Bar Chart~\cite{karduni_du_2020} offer improvements for analyzing two disparate ranges, their effectiveness diminishes with \omvs~\cite{borgo_order_2014}, as they require the accurate representations of multiple (\ie more than four) data ranges in the same view. 

    \begin{figure*}[htbp]
        \centering
        \includegraphics[width=0.85\textwidth]{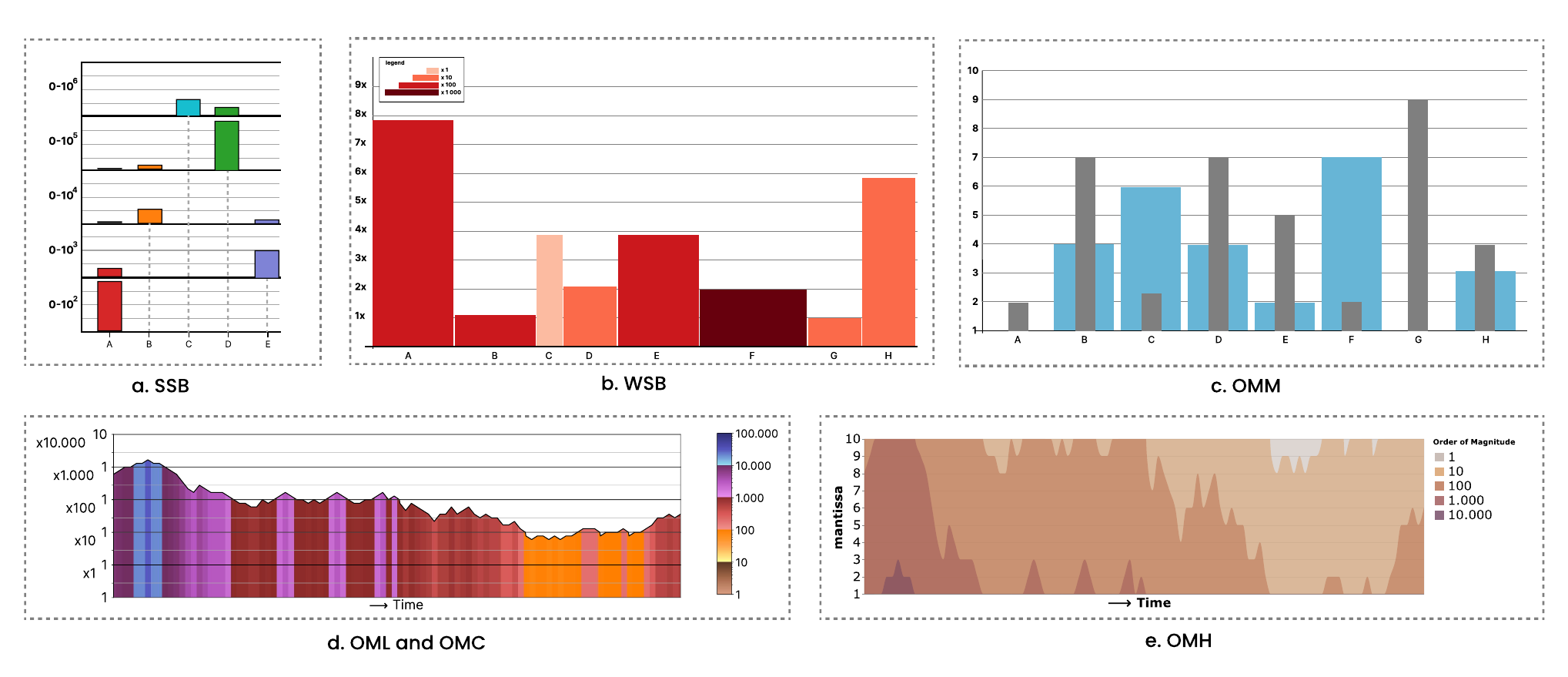}%
        \begin{subcaptiongroup}
        \phantomcaption\label{fig:SSB}%
        \phantomcaption\label{fig:WSB}%
        \phantomcaption\label{fig:OMM}%
        \phantomcaption\label{fig:OML}%
        \phantomcaption\label{fig:OMH}%
        \end{subcaptiongroup}%
        \caption{Mantissa-exponent visualizations: 
          \subref{fig:SSB} Scale-Stacked Bar Charts (\SSB)~\cite{hlawatsch_scale-stack_2013},
          \subref{fig:WSB} Width-Scale Bar Charts (\WSB)~\cite{hohn_width-scale_2020},
          \subref{fig:OMM} Order of Magnitude Markers (\OMM)~\cite{borgo_order_2014},
          \subref{fig:OML} Order of Magnitude Color and Line Chart (\OML)~\cite{braun_reclaiming_2023},
          \subref{fig:OMH} Order of Magnitude Horizon Graph (\OMH)~\cite{braun_reclaiming_2023}.}
        \label{fig:omv-visualizations}
        \Description{This image shows examples of related work on OMV visualizations that separate OMVs into mantissa and exponent. First (a) is the Scale-Stacked Bar Chart, which visualizes OMVs by stacking values at multiple scales, starting from zero and mapping numbers linearly. Second (b) is the Width-Scale Bar Chart, which uses the Y position for the mantissa and combines color and width to represent the exponent. The third (c) is the Order of Magnitude Markers, which visualizes exponents and mantissas as overlapping bars. The fourth (d) is the Order of Magnitude Line Chart, combining the positional encoding of the Scale-Stacked Bar Chart with an Order of Magnitude color scale. Finally, (e) is the Order of Magnitude Horizon Graph, which uses the Y position for the mantissa and color intensity for the exponent.}
    \end{figure*}

    \begin{table*}[ht]
        \centering
        \footnotesize
        \setlength{\arrayrulewidth}{0.2mm}
        \setlength{\tabcolsep}{6pt}
        \renewcommand{\arraystretch}{1.2}
        \begin{tabular}{|l|l|l|l|l|l|l|}
        \hline 
        \thead{Reference} & \thead{Data Type} & \thead{Mark\\ Type} & \thead{Range for\\Exponent} & \thead{Channel for Exponent} & \thead{Channel for Mantissa} & \thead{Tasks Evaluation} \\
        \hline  
        \SSB~\cite{hlawatsch_scale-stack_2013} & nominal & line & 5 ($10^0$–$10^4$) & Row & Y position & Value, Sort, Difference, Ratio \\
        \hline
        \WSB~\cite{hohn_width-scale_2020} & nominal & line & 4 ($10^0$–$10^3$) & Length and Color Intensity & Y position & Value, Sort, Ratio, Trend \\
        \hline  
        \OMM~\cite{borgo_order_2014} & nominal & line (2) & 5 ($10^1$–$10^{5}$) & Y position and Color Hue & Y position and Color Hue & Value, Sort, Ratio, Trend \\
        \hline  
        \OMC~\cite{braun_color_2022} & timeseries & area & 7 ($10^{-8}$–$10^{-2}$) & Color Hue & Color Intensity & Value, Sort, Difference \\
        \hline  
        \OML~\cite{braun_reclaiming_2023} & timeseries & area & 6 ($10^0$–$10^5$) & Y position and Color Hue & Y position and Color Intensity & Value, Sort, Difference, Trend \\
        \hline  
        \OMH~\cite{braun_reclaiming_2023} & timeseries & area & 5 ($10^0$–$10^4$) & Color Intensity & Y position & Value, Sort, Difference, Trend \\
        \hline 
        \end{tabular}
        \caption{Description of mantissa-exponent visualizations proposed in the literature.}
        \Description{This table summarizes all the related work in terms of the range of the exponent they covered, the encoding channels used for both mantissa and exponent, the type of the mark, and the task they evaluated OMVs.}
        \label{table:omv}
        \vspace{-2em}
    \end{table*}

    Logarithmic scales, capable of showcasing diverse ranges, facilitate some qualitative tasks with \omvs, like finding extrema~\cite{hohn_width-scale_2020}. Nonetheless, they pose difficulties for quantitative tasks, such as estimating differences or ratios between values of similar or close magnitudes~\cite{hlawatsch_scale-stack_2013}. Despite their prevalence in scientific contexts, they may lead to  misconceptions~\cite{kubina_slope_2023} and are less understandable for the general public, challenging perception and interpretation~\cite{ciccione_analyzing_2022, ryan_logarithmic_2020, landy_estimating_2013}. \changed{Research results in cognitive psychology~\cite{landy_estimating_2013,landy2017, ciccione_analyzing_2022, Engler2024} showcase a common misconception: individuals frequently interpret the spacing between major ticks on a logarithmic scale linearly. This segmented or piecewise linear interpretation becomes particularly pronounced when dealing with values exceeding one million, aligning with how large magnitudes are verbally and symbolically represented ~\cite{landy_estimating_2013,landy2017}.}

    \changed{When describing large numbers verbally, we use a hybrid notation comprising two components: a number ranging between 1 and 999 to indicate precision and a scale-word (\eg millions, billions) to represent magnitude. Similarly, when we write large numbers using scientific notation we use two components: the mantissa for precision and the exponent for magnitude. Both notations rely on a structured schema that separates a value into two parts, linked by a mathematical relationship. The exponent increases multiplicatively (e.g., powers of 10), while the mantissa increases additively, showing a linear progression within the range defined by two exponents. In the following sections we use the term \textit{continuity} to refer to this mathematical relationship between mantissa and exponent.}

    In the past decade, researchers have developed visualizations that separate \omvs into exponent and mantissa components, enhancing tasks like value retrieval, sorting, ratio estimation, difference estimation, and trend identification.
    Hlawatsch et al.~\cite{hlawatsch_scale-stack_2013} introduced the Scale-Stack Bar Chart (\textbf{\hypertarget{SSB}{SSB}}) that visualizes \omvs by stacking values at multiple scales, beginning from zero and mapping numbers linearly. 
    % Consistent with the segmented linear model approach, this visualization was evaluated as more effective than traditional linear and logarithmic bar charts for quantitative comparisons. 
    Borgo et al.~\cite{borgo_order_2014} visualized exponents and mantissas as overlapping bars, termed Order of Magnitude Markers (\textbf{\hypertarget{OMM}{OMM}}). Höhn et al.~\cite{hohn_width-scale_2020} presented Width-Scale Bar Charts (\textbf{\hypertarget{WSB}{WSB}}), using Y position for mantissa and color plus width for the exponent.
    % They compared \WSB with \SSB in a controlled experiment, showing that \WSB is more effective for value retrieval and \SSB for sort, ratio, and trend tasks. 
    Additionally, Braun et al.~\cite{braun_color_2022} used color in their Order of Magnitude Color (\textbf{\hypertarget{OMC}{OMC}}) scheme to map exponents to hue and mantissa to a color scale. Focusing on large-scale time series, Braun et al.~\cite{braun_reclaiming_2023} introduced the Order of Magnitude Horizon Graph (\textbf{\hypertarget{OMH}{OMH}}) and Line Chart (\textbf{\hypertarget{OML}{OML}}), with the first using Y position for mantissa and color intensity for exponents, and the second combining the positional encoding from \SSB and the \OMC color scale. \changed{Unlike \SSB, where values are shown per magnitude in stacked rows, \OML use ``linear scaling within orders of magnitude'' combining both exponent and mantissa in the Y position ~\cite{braun_reclaiming_2023}}. 
    % Unlike \SSB, where values are shown per magnitude in stacked rows, \OML combines both exponent and mantissa in the Y position, introducing a new visualization scale termed ``linear scaling within orders of magnitude''~\cite{braun_reclaiming_2023    We refine the definition of their scale and call it \emph{\EplusM} in \autoref{sec:EplusM}.}
    \autoref{fig:omv-visualizations} shows examples of these visualizations, and \autoref{table:omv} summarises existing literature. 

    The studies highlighted above have conclusively demonstrated some advantages of visualizations that separate \omvs into their mantissa and exponent components. This separation enhances analytical tasks' accuracy, confidence, and efficiency by offering a more effective representation of \omvs. However, these studies have explored only a subset of the potential combinations of marks and visual channels for encoding the mantissa and exponent (\autoref{table:omv}). Although this subset is promising, there is still a lack of a systematic approach in understanding the factors affecting the effectiveness of mantissa and exponent visualizations.
    To address this gap, our objective is to define a design space for visualizing mantissa and exponent, systematically explore it, document observed problems that may inform design guidelines, and identify effective encodings, enriching the corpus of \omv visualizations. We base our exploration and evaluation on the extensive literature on graphical perception and visualization effectiveness, summarized in the following section.

    \subsection{Graphical Perception and Visualization Effectiveness}\label{sec:perception}
    The concept of visualization effectiveness has evolved significantly since Bertin first introduced it~\cite{Bertin}. Building on Bertin's foundation, Cleveland and McGill~\cite{Cleveland84} conducted experiments to evaluate the ranking of visual channels for quantitative data, with a specific focus on the accuracy---how precisely a value can be identified~\cite{munzner2015visualization} of sorting and comparison tasks. Mackinlay~\cite{Mackinlay86} expanded this ranking to include both nominal and ordered data, thus broadening our understanding of visual channel effectiveness. Several subsequent experiments~\cite{Heer2010, Kim18, McColeman22} have empirically validated and refined the proposed effectiveness rankings of visual channels, revealing decoding factors, such as different tasks, that could influence them. Zeng and Battle~\cite{zeng2023recom}, through a systematic review of the literature on graphical perception, pointed out that position encodings are the most accurate channels across all data types. The \textit{accuracy} of other visual channels (\eg area, length, or color intensity) varies, influenced by factors like \textit{discriminability}---the ability to perceive distinct values for each attribute~\cite{munzner2015visualization}---, and \emph{separability}---the independence of channels from one another~\cite{munzner2015visualization}.
    
    Discriminability is affected by several factors, including the number of objects~\cite{HealeyE12}, the type and size of a mark~\cite{LiWM10, Szafir2018}, the background color~\cite{LiWM10}, visual attention limitations~\cite{HealeyE12}, and interference effects with other channels~\cite{Szafir2018}. Interference effects with other channels lead to separability issues. Separability allows for focused attention on one visual channel without interference from another~\cite{Garner1976}. Separable visual channels, such as position and color, show minimal cross-channel effects, causing no effectiveness issues when assigned to different attributes of a visualization. In contrast, when the encoding of one channel affects the perception of another, the channels are called \emph{integral}~\cite{Garner1976}. Integral visual channel pairs include color hue and color intensity~\cite{Gottwald1975, Callagha1984}, height and width~\cite{krantz_similarity_1975, macmillan_mean-integral_1998, ganel_visual_2003, Tanzer2013}, shape and size~\cite{Smart2019}, shape and color intensity~\cite{Smart2019}, size and color intensity~\cite{Szafir2018}, and size and position~\cite{Kim18}. The redundant combination of integral visual channels for the same attribute facilitates decoding. However, encoding two unrelated attributes using two integral channels can lead to interference, hindering identification and value decoding.~\cite{Algom2016}.  

    Visualization researchers have leveraged this knowledge to develop quality metrics~\cite{Behrisch2018}, heuristics~\cite{Zuk2006_heuristics}, design guidelines~\cite{borgo_guidelines_2013}, and recommendation systems~\cite{Drako2019}, which are used to design visualizations and evaluate their effectiveness. Although some of these heuristics lay the groundwork for our exploration, they are not yet sufficient to determine whether a visualization that separates \omvs into mantissa and exponent is effective. In such visualizations, the perceptual task of magnitude estimation depends on the combination of two values---mantissa and exponent---rather than a single one. As a result, well-established low-level perceptual quality metrics for magnitude estimation may not fully apply in this context. Potential interference between the visual encodings of mantissa and exponent could complicate the estimation process. Additionally, disregarding familiar mental models of numbers, such as the continuity between mantissa and exponent that mirrors verbal number expression, may hinder accurate comparisons. 
    
    This gap underscores the necessity for systematic exploration and evaluation of how mantissa and exponent are visually encoded and how these encodings are decoded perceptually and cognitively in various tasks. By identifying and documenting encoding and decoding challenges, our research aims to enrich the graphical perception literature by introducing guidelines that specifically support low-level perception and cognition tasks such as magnitude estimation and quantitative comparisons for visualizations that separate \omvs into mantissa and exponent.

\section{Design space}
 In this section, we describe a design space for visualizing tabular data with one quantitative \omv attribute, separated into mantissa and exponent. We follow a constraint-based approach~\cite{Biskjaer2014}, setting specific boundaries and refining the space by insights gained from the literature, related to: 1) the \changed{\gog~\cite{GrammarOfGraphics}} 2) studies on graphical perception~\cite{Algom2016, QuadriR22} and visualization effectiveness~\cite{Bertin, zeng2023recom, Drako2019}, and 3) the literature on \added{perception~\cite{landy_estimating_2013, landy2017, ciccione_analyzing_2022} and } visualization~\cite{braun_reclaiming_2023, borgo_order_2014, hohn_width-scale_2020}  of large value ranges. 
 
 Considering the combinatorial nature of all the choices a designer can make to visualize mantissa and exponent---such as choosing between different marks, visual channels, layouts, developing novel scales, employing redundant encoding, and adding design embellishments--- describing, exploring, and evaluating the entire design space of all candidate solutions is infeasible. Therefore, we initially focused on a well-structured\added{~\cite{Simon73}} subset of the design space, including visualizations with high data-to-ink ratio (no redundancy, single mark, no embellishments-with exception grid lines and tick marks), supported by current implementations of the \changed{\gog}, such as Vega-Lite~\cite{Vega-Lite} and Plot~\cite{Plot}. This approach allowed us to observe the effectiveness of assignments for mantissa and exponent while minimizing interference from external design factors and to systematically generate a large number of visualizations. 

We start by establishing the boundaries of our design space, proceed to describe its dimensions---how each visualization within this space is described---then we define the \EplusM scale, and conclude by outlining the constraints implemented to refine the design space, with the goal of improving visualization effectiveness.

 \subsection{Boundaries}
 Our design space's boundaries are shaped by the objectives of our research and insights gained from the literature review. We concentrate on tabular datasets with a single \omv attribute, excluding types such as networks, fields, and geometries. Secondly, we focus on the Cartesian coordinate system, excluding polar or parallel layouts. Aiming to enhance the effectiveness of \omv visualizations in overview and design visualizations that are applicable to both printed and digital media, we limit our scope to static visualization designs and exclude interactive elements. Moreover, our design space does not entail data transformations; these are handled during dataset creation. Our analysis explicitly targets visualizations that employ a single mark, and consequently, we are removing the Order of Magnitude Markers~\cite{borgo_order_2014} from our design space. 

  %  \begin{figure*}
  %   \includegraphics[width=0.9\textwidth]{design-space}
  %   \caption{ Our design space encompasses three dimensions: MARKS (green), DATA (blue), and VISUAL CHANNELS (red). The image illustrates an example of using our design space as an interactive table, where a mark is selected, and visual channels are assigned to data attributes. Grey cells are invalid, according to the visualization literature. After checking for integrity constraints, a visualization is generated to perform the tasks. Note the $\bigoplus$ signs (top left); enable the assignment of the \EplusM scale to positions.}
  %   \label{fig:design-space}
  %   \Description{The image shows our design space and the defined dimensions. Our design space encompasses three dimensions: MARKS (green), DATA (blue), and VISUAL CHANNELS (red). The image illustrates an example of using our design space as an interactive table, where a mark is selected, and visual channels are assigned to data attributes. Grey cells are invalid, according to the visualization literature. After checking for integrity constraints, a visualization is generated to perform the tasks.}
  % \end{figure*}
 
\subsection{Dimensions}
After setting the design space boundaries, we define the dimensions that specify each visualization. \gog implementations~\cite{Plot, ggplot} describe a visualization as the combination of data types, marks, visual channels, facets, scales, transforms, and projections. Considering the boundaries established, we describe each visualization design by a combination of the following elements: data, marks, and visual channels, considering faceting as part of visual channels~\cite{Vega-Lite, zeng2023recom, Drako2019}.  
    \autoref{fig:design-space} shows the dimensions of our design space.
    
    \paragraph{Data}
    In this context, ``data'' pertains to the type and range of attributes. We consider \omv as a specific type of quantitative data attribute that is separated into two quantitative attributes: \textbf{mantissa} and \textbf{exponent}, as defined in the background section. We explore visualizations that display the mantissa, the exponent, and an additional attribute type (we call \emph{\otherattr}) among \textbf{nominal}, \textbf{ordinal}, \textbf{temporal}, and \textbf{quantitative}. For the exploration of the design space we created datasets with one-to-one relationship between the \omv and the \otherattr.
    % The range of these additional data attributes, and therefore the cardinality of the datasets chosen for our exploration, was seven, small enough to remain discriminable with all the visual channels~\cite{Healey96}. 
    \autoref{fig:design-space} displays the DATA (blue) part as rows on the top left of the table.

    \paragraph{Marks}
    Marks are defined based on their spatial dimension requirements: \textbf{points} represent zero-dimensional marks, \textbf{lines} correspond to one-dimensional marks, and \textbf{areas} are considered two-dimensional marks~\cite{Bertin, munzner2015visualization}. Within the framework of the \gog implemented by Plot, we represent points either with the Dot or with the Rule mark---when the visual channel of Length is chosen---, for lines, we use the Bar mark for nominal and ordinal data, the Rect mark for time, and last, for areas, we use the Area mark. \autoref{fig:design-space} displays the MARKS (green) part as rows on the left of the table.

    \paragraph{Visual channels}
    Visual channels (VC) are ways to control the appearance of marks~\cite{munzner2015visualization}. In our design space, we selected channels based on insights from the literature on visualization effectiveness~\cite{zeng2023recom, Drako2019}. We focus on the following nine visual channels: X position (\posx), Y position (\posy), \row, Column (\col), \len, \area, Color Intensity (\colint), Color Hue (\colhue), and \shape. Length refers to both width and height, depending on the orientation of the line and the choice of the mark. \row and \col are channels derived from the faceting operation~\cite{Vega-Lite, zeng2023recom, Drako2019}. \autoref{fig:design-space} displays the VISUAL CHANNELS (red) as columns on the top of the table.
    \[
    \VC = \{ \posx, \posy, \row, \col, \len, \area, \colint, \colhue, \shape \}    
    \]

   \paragraph{Description of Visualizations}
    Considering the dimensions above, we describe a visualization that belongs to our design space as a list of four \emph{assignments} forming a \emph{visualization configuration}:\\ Mark, Exponent $\mapsto \VC$, Mantissa $\mapsto \VC$, Attribute-2 $\mapsto \VC$.
    
    \autoref{fig:design-space} displays the combination on the bottom of the generated visualization as:
    \begin{figure}[ht]
        \centering
        \includegraphics[width=0.6\linewidth]{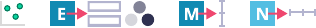}
         \caption{Point, Exp. $\mapsto \row$ and \colint, Mant. $\mapsto \posy$, Nominal $\mapsto \posx$}
        \Description{The image describes visually how a visualization is described in our design space. Giving the following example: Point, Exp. ↦ Row and Intensity, Mant. ↦ PosY, Nominal ↦ → PosX }
        \label{fig:description-of-visualization}
    \end{figure} 

    \paragraph{Scales} According to the \gog, a visualization scale is a function that transforms attribute values into visual channel values, such as position, length, or color intensity. In our design space, scales are handled by Plot and are determined automatically by the attribute type. Band scales are utilized for nominal attributes, ordinal scales for ordinal attributes, and linear scales for quantitative attributes. 
    To simplify interpretation and eliminate the need for calculations, we implement a tick transformation representing the exponent value as $10^e$. For instance, if the exponent is 3, the tick label displays 1,000 ($10^3$). Given that scales are selected based on the data attribute type, they are not discussed within our design space, with the exception of \EplusM (see below).

    \subsection{\EplusM Scale}\label{sec:EplusM}
    As mentioned in the background, studies in cognitive psychology~\cite{landy_estimating_2013} suggest that people perceive \omvs following a segmented linear model, with magnitudes distributed uniformly as in logarithmic scales and numbers within these magnitudes being linearly interpolated. We believe the scale used by Braun et al.~\cite{braun_reclaiming_2023}, and inspired by the positional encoding of \SSB~\cite{hlawatsch_scale-stack_2013}, \changed{aligns closely with this segmented linear model}. \added{However, while Braun et al.~\cite{braun_reclaiming_2023} encode exponent and mantissa  continuously on a single axis, they still display the two components in y-axis separately (see \autoref{fig:OML}), requiring users to multiply them to derive the corresponding value.} 
    
    \changed{We integrate the mantissa and exponent into a single positional scale, termed \EplusM (exponent plus mantissa) and refine its definition to facilitate experimental evaluation and encourage broader adoption in \omv visualizations. \EplusM is a piecewise linear scale that maps a value (v) to visual elements using the formula: $s(v) = e(v) + (m(v)-1)/ 9$, where \(e(v)\) is the exponent and \(m(v) \in [1, 10)\) is the mantissa. This scale uses the exponents as primary units and fills the $[0,1)$ interval between consecutive exponents, by rescaling the mantissa to $(m-1)/9$. When applied to positional encoding, the axis units (major ticks) represent the values of the exponents, and the subunits (minor ticks) correspond to the mantissa values. A tick transform is necessary to show the original value \(v\) corresponding to each \(s(v)\). We recommend displaying the midpoint between two exponents to highlight the linearity of the scale and illustrate the segmented linear model.} \added{The supplemental material includes an example of \EplusM implementation.} In our design space, this scale is described by assigning both the exponent and mantissa to the same positional channel (\eg Exponent $\mapsto \posy$, Mantissa $\mapsto \posy$).

\subsection{Constraints}
    To ensure the effectiveness and expressiveness of the visualizations in our design space, we apply constraints related to the boundaries we have set, the expressiveness of marks and visual channels, and the interference of different channels. We consider a visualization \emph{viable} only if all constraints are satisfied. Our approach is inspired by the literature on recommendation systems for visualization~\cite{Drako2019}, where integrity constraints are used to encode expressiveness and restrictions to the dimensions of a visualization specification. 
    
    \paragraph{Constraints based on layout} 
    As we limit our design space to the Cartesian system, we include only combinations that contain at least one position channel (\posx or \posy).

    \paragraph{Constraints for the expressiveness of channels} 
    To ensure the expressiveness of each assignment~\cite{munzner2015visualization}, we use \colhue and \shape only for nominal data, \colint for ordinal and quantitative data, and \len and \area only for quantitative data. For the \row and \col channels, we assign only the exponent, as mantissa is a quantitative attribute with continuous range, and 
    % the other attributes 
    \otherattr has a one-to-one relationship with the \omv, leading to visualizations similar to those using a positional channel. An exception to this constraint is our decision to retain visualizations that include both row and column for the exponent and \otherattr as a nominal attribute, as they result in novel visualizations.
    
    \paragraph{Constraints based on mark type} 
    For each type of mark, we impose constraints in accordance with its defined dimensions. The \shape and \area channels can be used only for point marks, and the \len channel is not feasible for area marks. We also apply constraints in relation to the dataset. We use area marks only for temporal data, and we exclude lines for quantitative data.

    \paragraph{Constraints based on interference with identity channels} 
    Channels like \colhue and \shape serve identification purposes. Combining \colhue (\eg red) with \colint within the same mark generates a new color (\eg pink). Similarly, merging \shape with \len can alter the shape (\eg modifying a circle into an ellipse). These combinations modify the identity of the primary channel, and thus, we do not consider combining \colhue with \colint and \shape with \len.

\section{Generation of Visualizations} 
 \begin{figure*}[ht]
   \begin{subcaptiongroup}
    \includegraphics[width=0.9\textwidth]{img/examples-design-space.pdf}
    \phantomcaption\label{fig:vNominal}
    \phantomcaption\label{fig:vQuantitative}
    \phantomcaption\label{fig:vTemporal}
    \phantomcaption\label{fig:vOrdinal}
    \end{subcaptiongroup}
    \caption{\added{Two examples of visualizations per dataset: 
      \subref{fig:vNominal} Nominal,
      \subref{fig:vQuantitative} Quantitative,
      \subref{fig:vTemporal} Temporal, and
      \subref{fig:vOrdinal} Ordinal.
      All the visualizations were generated using our OMVis Tool. At the top, visualizations are described according to the design space dimensions (see \autoref{fig:description-of-visualization}).}}
    \label{fig:examples-visualizations}
  \vspace{-1.2em}
    \Description{The image illustrates two examples of visualizations per dataset. At the top left, there are two visualizations of the Nominal dataset. The first uses a point mark, encoding the mantissa with PosY, the exponent with Area, and the nominal attribute with Shape. The second uses a line mark, encoding the exponent with PosY, the mantissa with Length, and the nominal attribute with PosX. At the top right, there are two visualizations of the Quantitative dataset. The first uses a point mark, encoding the mantissa with PosY, the exponent with Area, and the nominal attribute with Shape. The second uses a line mark, encoding the exponent with PosY, the mantissa with Length, and the nominal attribute with PosX. At the bottom left, there are two visualizations of the Temporal dataset. The first uses an area mark, encoding the mantissa with PosY, the exponent with Intensity, and the temporal attribute with PosX. The second uses a line mark, encoding both the exponent and the mantissa with PosY (EplusM scale), and the nominal attribute with PosX. At the bottom right, there are two visualizations of the Ordinal dataset. The first uses a point mark, encoding the mantissa with PosY, the exponent with Row, and the ordinal attribute with Column. The second uses a point mark, encoding the exponent with Area, the mantissa with Intensity, and the nominal attribute with PosX. All the visualizations were generated using the “Order of Magnitudes Tool” (see Section 4.1). At the top, visualizations are described according to the design space dimensions (see Figure 4).}
  \end{figure*}
We developed a tool that enabled us to systematically generate and explore all possible assignments of visual channels to mantissa and exponent across marks and data attribute types. \added{\autoref{fig:examples-visualizations} shows examples of the generated visualizations.}

 \subsection{Order of Magnitudes Tool}\label{sec:tool}
To facilitate the exploration of the design space, we implemented a web-based open-source tool, OMVis, written in JavaScript. Our tool includes two main components: 1) a menu for the interactive exploration of all possible combinations and 2) the resulting visualization. Within the menu, users can upload or select a dataset from our database. They then choose an \omv to separate into its mantissa and exponent parts, followed by the selection of additional attributes (nominal, ordinal, time, or quantitative). Subsequently, users decide on a mark (point, line, or area) and assign visual channels to the mantissa, exponent, and selected attributes. We apply the constraints by disabling the corresponding choices; for example, quantitative visual channels such as \len are not applicable to nominal attributes. A visualization is generated based on these selections.  We use Plot to create the visualizations. The user can export the produced visualization either as an image or as JavaScript code. More details are provided in the supplemental material.

\subsection{Datasets}
We used the ``Trending YouTube Video Statistics'' dataset from Kaggle~\cite{Kaggle}, which comprises daily metrics (views, likes, comments) for the top trending YouTube videos across various categories. We selected this dataset because it contains \omvs (e.g., \# of views) and includes diverse attributes such as nominal (e.g., video categories), ordinal (e.g., 7-point Likert scale sentiments), temporal (e.g., days), and quantitative (likes, comments). For the design space exploration, we created four smaller datasets (maximum cardinality = 18), each featuring views (\omv) and another attribute: nominal, ordinal, temporal, or quantitative. The exponent range is from five to eight. The 
% original dataset and the Jupyter notebook for creating the four 
datasets used in our exploration are available in the supplemental material.
    
\subsection{Visualizations Generation Process}
We systematically generated all visualizations by first calculating every combination of the nine visual channels ($n=9$) in our design space with our data attributes ($k=3$) \{\textit{exponent}, \textit{mantissa}, \textit{\otherattr}\}, where \textit{\otherattr}  represents a nominal, ordinal, temporal, or quantitative attribute. For our exploration, we considered only one-to-one assignments, where one data attribute is encoded with one visual channel. This constitutes a permutation problem since the order of assignment matters (\ie which visual channel is assigned to which data attribute). The formula for permutations of $n$ items taken $k$ at a time is given by:  $P(n, k) = \frac{n!}{(n-k)!}$ so, in our case, $P(9,3)=504$ visualizations. 

An exception to the one-to-one rule involves the visual channels of \posx and \posy, where both the mantissa and exponent can be assigned to the same visual channel, resulting in the scale denoted as \emph{\EplusM}. The total number of combinations for the \emph{\EplusM} scale for both \posx and \posy is $2 \times P(8,1) = 16$ visualizations.
    
This method was uniformly applied across all mark types \{\textit{point}, \textit{line}, \textit{area}\} and data attribute types for \otherattr \{\textit{nominal}, \textit{ordinal}, \textit{time}, \textit{quantitative}\}, resulting in the total number of combinations before imposing any constraints being equal to:
\[ %begin{equation}\label{eq:total}
    3 \times 4 \times (504 + 16) = 6,240
\] %end{equation}

The result after the application of constraints mentioned in the previous section was 336 viable combinations. We used our tool to generate all the viable visualizations and exported them as images. For details about the creation of the combinations and the applications of constraints, see the program \texttt{combinations.js} in the supplemental material.

\section{Qualitative Evaluation} 
We qualitatively evaluated all generated visualizations using a systematic, inspection-based method inspired by the heuristic walkthrough technique~\cite{Sears_1997}. This two-pass evaluation approach guides evaluators with a set of heuristics, a prioritized list of user tasks, and ``thought-focusing'' questions~\cite{Sears_1997}. During each pass, evaluators document problems, assign individual ratings, and later compare and discuss the identified issues.

Building on this technique, we conducted two evaluation sessions. The first, called \textbf{Encoding Walkthrough}, was a group evaluation where we displayed all visualizations on a wall-sized screen and documented issues related to the encoding of \omvs, guided by visualization-specific questions and heuristics. The second, called \textbf{Decoding Walkthrough}, involved individual evaluations where each evaluator completed specific tasks, rated the visualizations, and noted any problems affecting the effectiveness of decoding the visualized \omvs.

Finally, we entered a collaborative phase to discuss and compare our ratings and observations. At the end of this process, we classified each visualization into four categories based on the identified problems and presented the insights gained as a set of design guidelines for visualizing mantissa and exponent.
% The goal of this stage was not to quantitatively compare the effectiveness of visualizations but to observe as many problems as possible and identify a set of encodings for mantissa and exponent that both follow heuristics for magnitude estimation and facilitate quantitative comparisons.

\subsection{Evaluators}
The group of evaluators (henceforth referred to as ``we'') who conducted the qualitative evaluation consisted of four experts in data visualization and/or human-computer interaction, all of whom are co-authors of this paper: one doctoral student, one practitioner and two faculty members. Each evaluator has a range of 5-30 years (mean=14 years) of research and practical experience in data visualization and/or human-computer interaction. Our coding methodology adheres to the tradition of expert-based evaluation of user interfaces, such as heuristic evaluation and cognitive walkthrough, where it has been shown that a small number of experts can reliably identify most problems~\cite{Mankoff2003, Tory2005}. 

% Rather than relying exclusively on the personal judgment of the evaluators, we informed and guided the evaluation sessions based on questions and heuristics derived from the visualization literature~\cite{munzner2015visualization}.
% 

\subsection{Encoding Walkthrough}
The focus of this session was to identify problems relating to the visual encoding of \omvs. This evaluation was guided by visualization heuristics related to the effectiveness of visual channels~\cite{munzner2015visualization, Zuk2006_heuristics, Tarrell2014} with a focus on aspects that could impact magnitude estimation (see \autoref{sec:perception}). We displayed all the visualizations on a wall-sized screen (6m wide and 2m high) and conducted a ``walkthrough'' of each. For every visualization, we aimed to answer the following heuristics-inspired thought-focusing questions:  
\begin{itemize}
    \item \textbf{Accuracy:} How precisely can we tell the difference between the different values of each data attribute?
    \item \textbf{Discriminability:} How many unique values can we perceive for each attribute? 
    \item \textbf{Separability:} Do the channels interfere with each other?
    \item \textbf{Grouping:} Can we identify values belonging to the same exponent? 
    \item \textbf{Continuity:} Do the channels used for the mantissa and exponent convey the numerical relationship between the two attributes?
\end{itemize}
We labeled each visualization based on whether or not we observed issues related to the previously defined questions regarding the mantissa, exponent, or the \otherattr, while also describing, the specific identified problems. For example, if there was a problem related to the separability between channels, we labeled separability with ``no'', providing more details about the problem's origin (\eg interference effect between \colint used for \otherattr and \area for exponent). Additionally, we evaluated the suitability of the displayed encodings for visualizing \omvs, responding with ``yes,'' ``no,'' or ``maybe.'' 
Using the Miro tool~\cite{Miro}, we then grouped the 336 visualizations according to their suitability labels (yes, maybe, no) and identified patterns related to the effectiveness of different visual encodings, taking into account the problems observed. \autoref{tab:problems} shows a summary of the encoding-identified problems.

We found that several identified issues were caused by the high cardinality (18) of the chosen datasets, impacting the readability of visualizations and hiding potential problems. As a result, for the second stage of our evaluation process, we reduced the cardinality to 7---small enough to ensure discriminability across all channels---by selecting representative values from each order of magnitude. Additionally, we observed that all symmetrical visualizations shared similar strengths and weaknesses. To streamline the evaluation, we removed the duplicated symmetrical visualizations, reducing the total number of visualizations from 336 to 168. We prioritized symmetrical combinations involving both position and faceting, opting to retain \posy and \row over \posx and \col~\cite{Conati2014, Kim18}.

\subsection{Decoding Walkthrough}
The focus of this session was to identify issues affecting the decoding of visualized \omvs, specifically in relation to three tasks: value retrieval, difference estimation, and ratio estimation. Based on the literature on visualizing \omvs~\cite{hohn_width-scale_2020, hlawatsch_scale-stack_2013, braun_reclaiming_2023}, we considered three conditions for difference and ratio estimation based on the relationship between the compared values: within the same, neighboring, or distant orders of magnitude (\ie the exponent of the values differ by at least two).

To conduct the evaluation, we developed an online tool. We assigned 84 visualizations to each evaluator, and therefore each visualization was evaluated by a pair of evaluators. 
% Each evaluator was paired with the three others, resulting in $C\binom{4}{2}$ = 6 pairs. Each pair evaluated 28 visualizations. 
Each evaluator performed the defined tasks and provided ratings on perceived accuracy for both the exponent and mantissa, the effort required to complete the task, the confidence in the provided answers, and the overall suitability of the visualization for \omvs. Additionally, evaluators could leave comments on any task-related issues they observed. Following the individual coding phase, we held a collaborative session to discuss and review all ratings. Inspired by collaborative coding methods in thematic analysis~\cite{braun2021Thematic,Richards2018CollabQA}, this collaborative session aimed to compile similar identified problems and reach a common agreement for all the visualizations.

\subsection{Results}
    \begin{figure*}
        \setlength{\belowcaptionskip}{-1em}
        \centering
        \includegraphics[height=5cm]{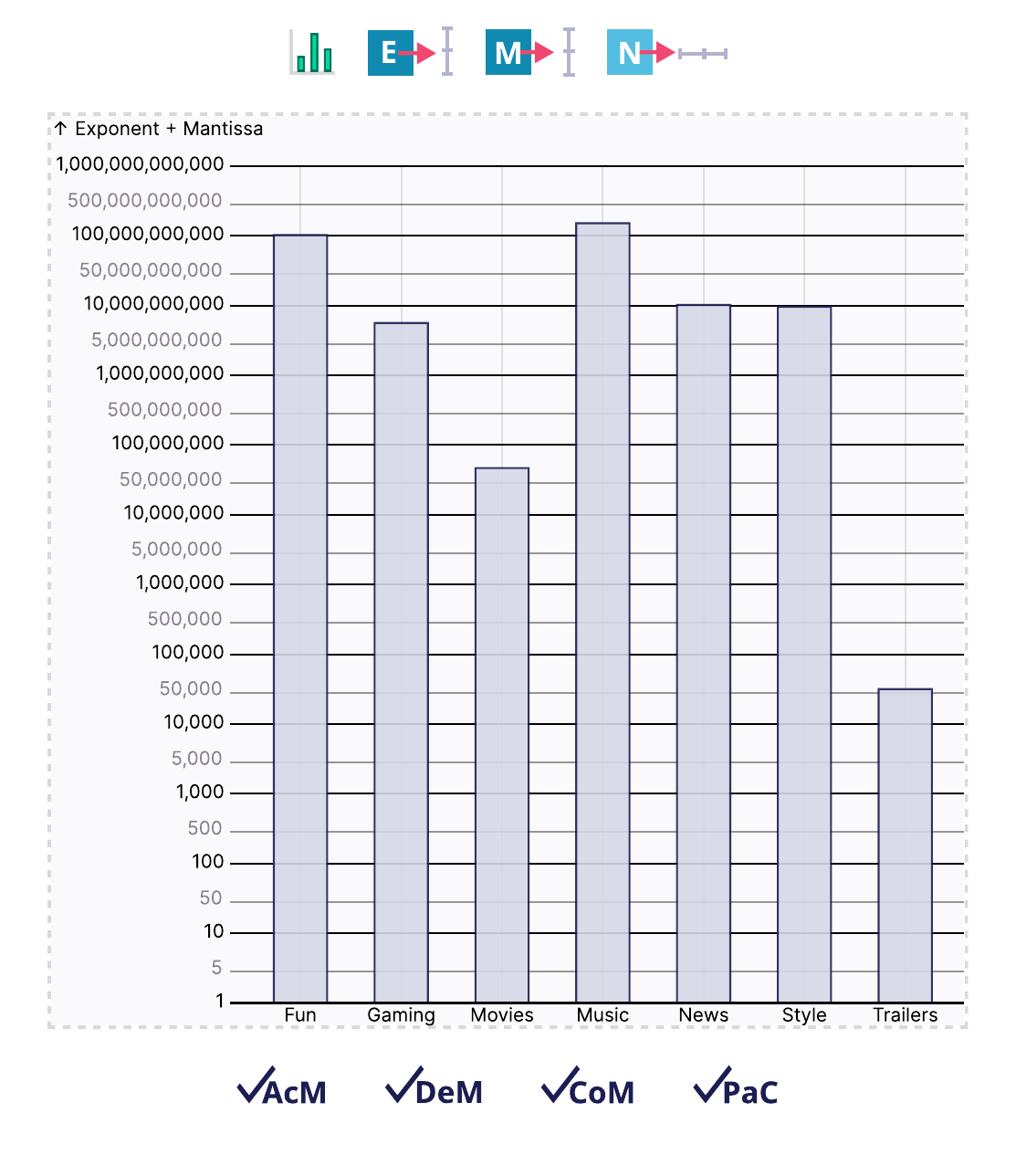}
        \includegraphics[height=5cm]{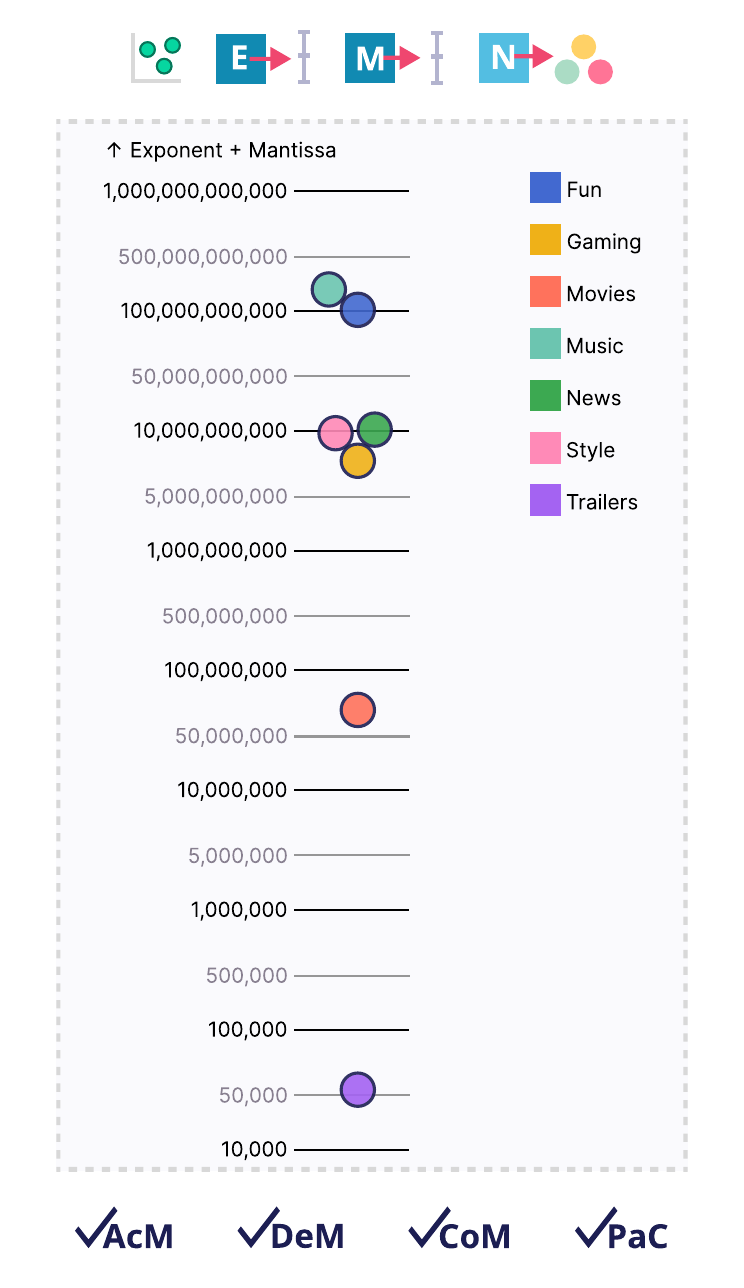}
        \includegraphics[height=5cm]{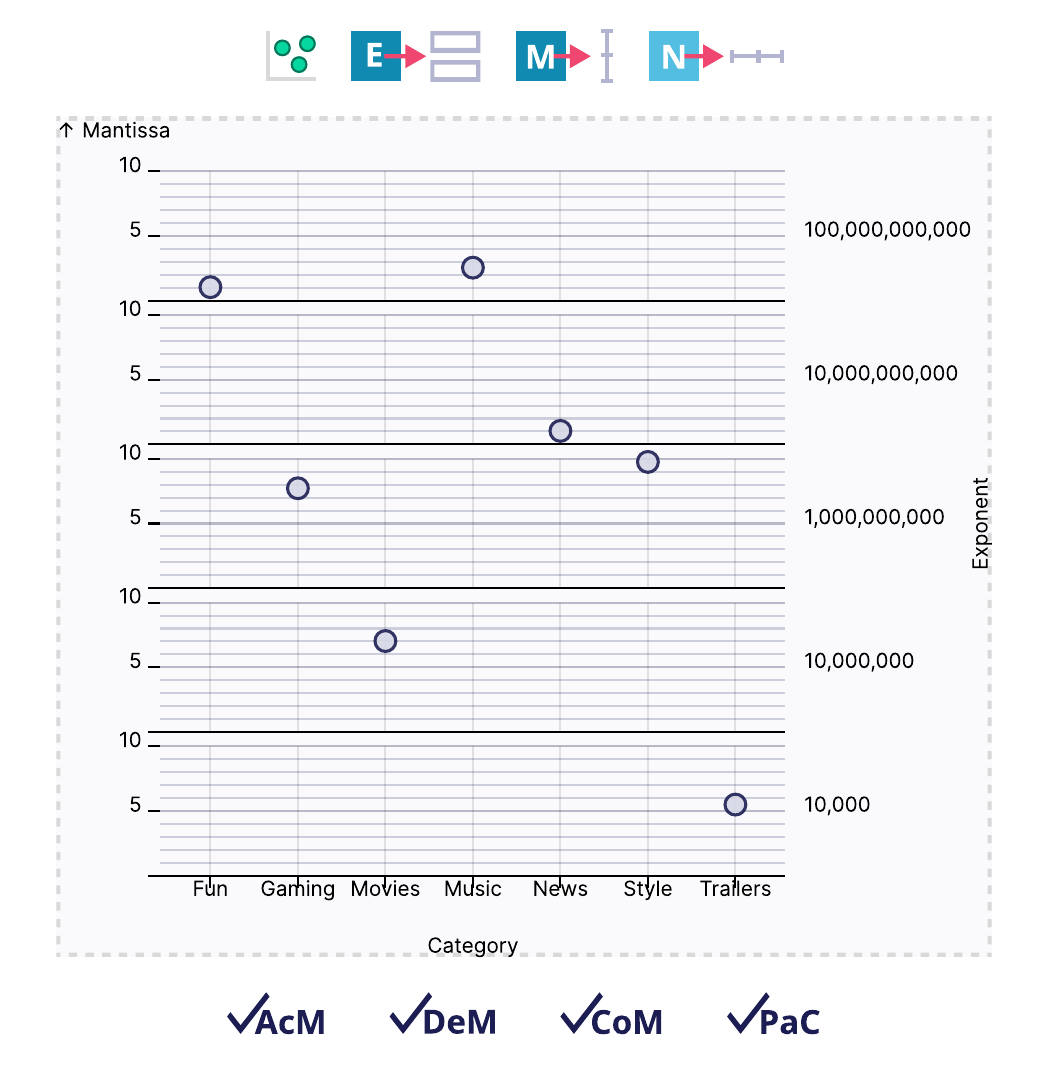}
        \caption{Sample of ``Effective assignments'' combinations (3/25) \added{. At the top, visualizations are described according to the design space (see \autoref{fig:description-of-visualization}), while at the bottom are abbreviations for the guidelines (see \autoref{sec:guidelines}) each visualization follows (\cmark). This sample of visualizations use positional channels for both the mantissa and exponent: the first two demonstrate the EplusM scale on the y-axis, and the third provides an example of Facet (the exponent is encoded with Row, and the mantissa with PosY).}}\label{fig:goodcomb}
        \Description{The image shows three of the visualizations we identified as "effective assignments" based on our qualitative evaluation. The first visualization uses the EplusM scale on the y-axis, with categories encoded by x-position using a line mark. The second visualization employs the EplusM scale with categories encoded by color hue using a point mark. The final visualization encodes the exponent using rows, the mantissa using the x-position, and the categories also using the x-position, with a point mark. }

    \end{figure*}

    \begin{table}[ht]
        \centering
        % \resizebox{\textwidth}{!}{ % Ensure table fits within the width of the page
        {
        \begin{tabular}{|p{0.4\linewidth}|p{0.5\linewidth}|
        %p{2cm}|p{2.1cm}|p{2cm}|p{2cm}|
        }
        \hline
        \textbf{Encoding Problems} & \textbf{Decoding Problems} 
        %& \textbf{1.Effective\newline assignments} & \textbf{2.Value retrieval, but no comparisons} & \textbf{3.Magnitude overview, but no detail} & \textbf{4.Impractical for \omvs} 
        \\ 
        \hline
        Low accuracy/ discriminability for exponent & Increased effort to perform the tasks, potentially leading to an order of magnitude error. %& no & no & no & yes 
        \\ 
        \hline
        Low accuracy/ discriminability for mantissa & Decreased confidence for value estimation and increased effort for quantitative comparisons. %& no & no & yes & maybe 
        \\ 
        \hline
        No continuity between\newline mantissa and exponent & Increased effort for quantitative comparisons; the user needs to retrieve and multiply mantissa and exponent to derive the value. % & no & yes & maybe & yes 
        \\ 
        \hline
        Low accuracy/ discriminability for the other attribute & Reduced readability, increased effort to perform the tasks. %& no & maybe & maybe & maybe 
        \\ 
        \hline
        Interference effects & Reduced readability, increased effort to derive the value. %& no & no & no & yes 
        \\ 
        \hline
        No grouping for exponent & Increased effort for quantitative comparisons. %& no & no & no & yes 
        \\ 
        \hline
        Encoding constraints\newline based on data cardinality & Reduced readability. Constraints in visualization size and visual channels that can be used for the other attribute. %& no & yes & yes & yes 
        \\ 
        \hline
        Encoding constraints\newline based on mark & Constraints in visual channels that can be used due to interference effects with the mark. Potential need for redundant encoding to increase effectiveness. %& maybe & yes & yes & yes 
        \\ 
        \hline
        Encoding constraints\newline based on data characteristics. & Constraints in visual channels that can be used. Need for design embellishments to address and communicate potential issues to users. %& maybe & yes & yes & yes 
        \\ 
        \hline
        \end{tabular}
        }
        \caption{Encoding and corresponding decoding problems that we observed during the qualitative evaluation.} 
        \Description{The table describes the encoding and decoding problems identified during the qualitative evaluation, based on which we created groups. The first column outlines the encoding problems, while the second addresses the corresponding decoding issues. The encoding problems include low accuracy and discriminability for mantissa and exponent, separability issues between channels, lack of continuity between mantissa and exponent, absence of grouping, and data or mark-related issues. The decoding problems are associated with increased effort, decreased confidence, and reduced readability.}
        \label{tab:problems}
        \vspace{-1em}
    \end{table}

    \begin{figure*}
    \centering
    \captionsetup{justification=centering} % Set caption justification to centering
        \subcaptionbox{``Value retrieval,\newline but no comparisons''
            \label{fig:bad1}}{
            \includegraphics[height=5cm]{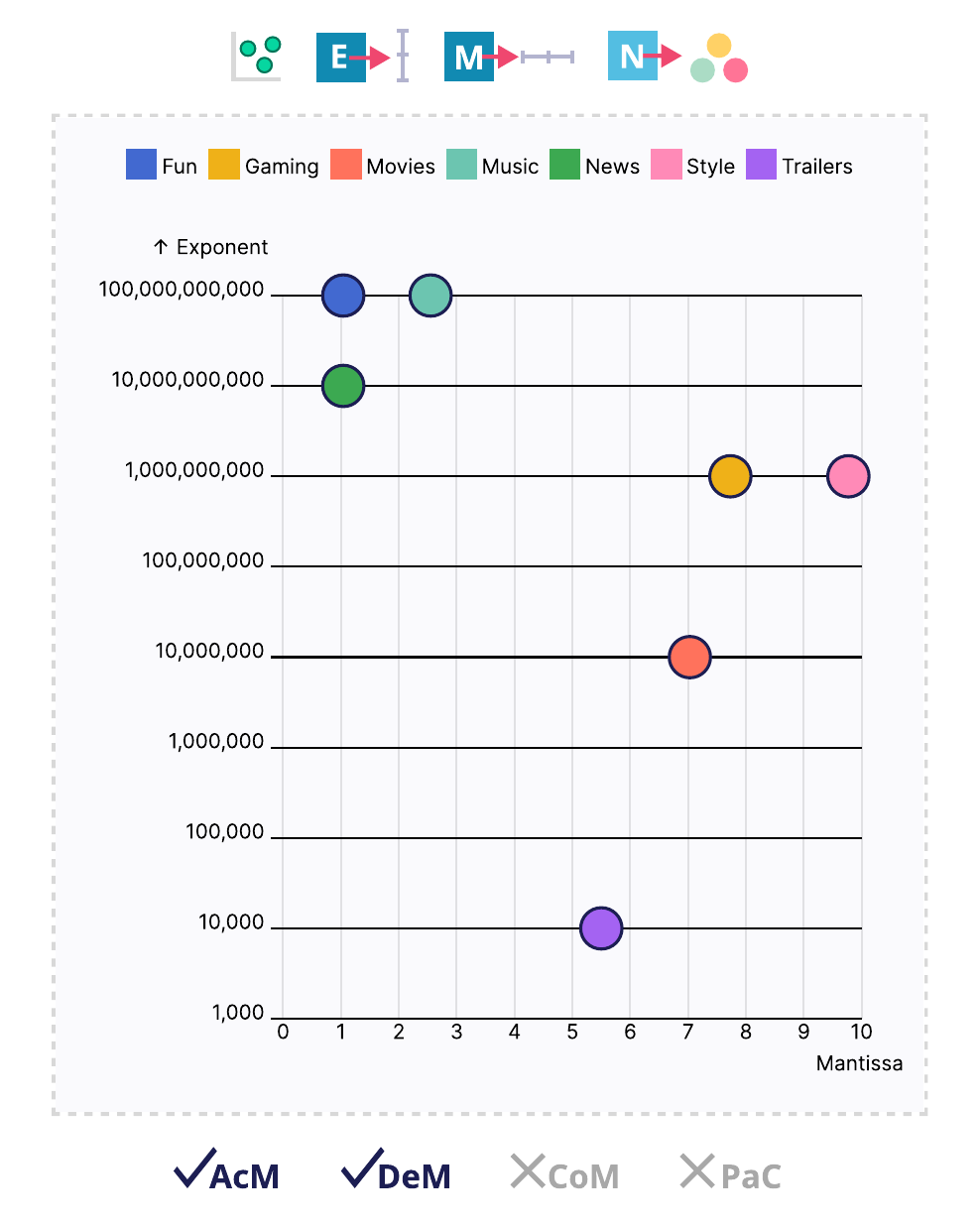}}
        \hspace{2em} % Add some horizontal space between images
        \subcaptionbox{``Magnitude overview,\newline but no detail''
            \label{fig:bad2}}{
            \includegraphics[height=5cm]{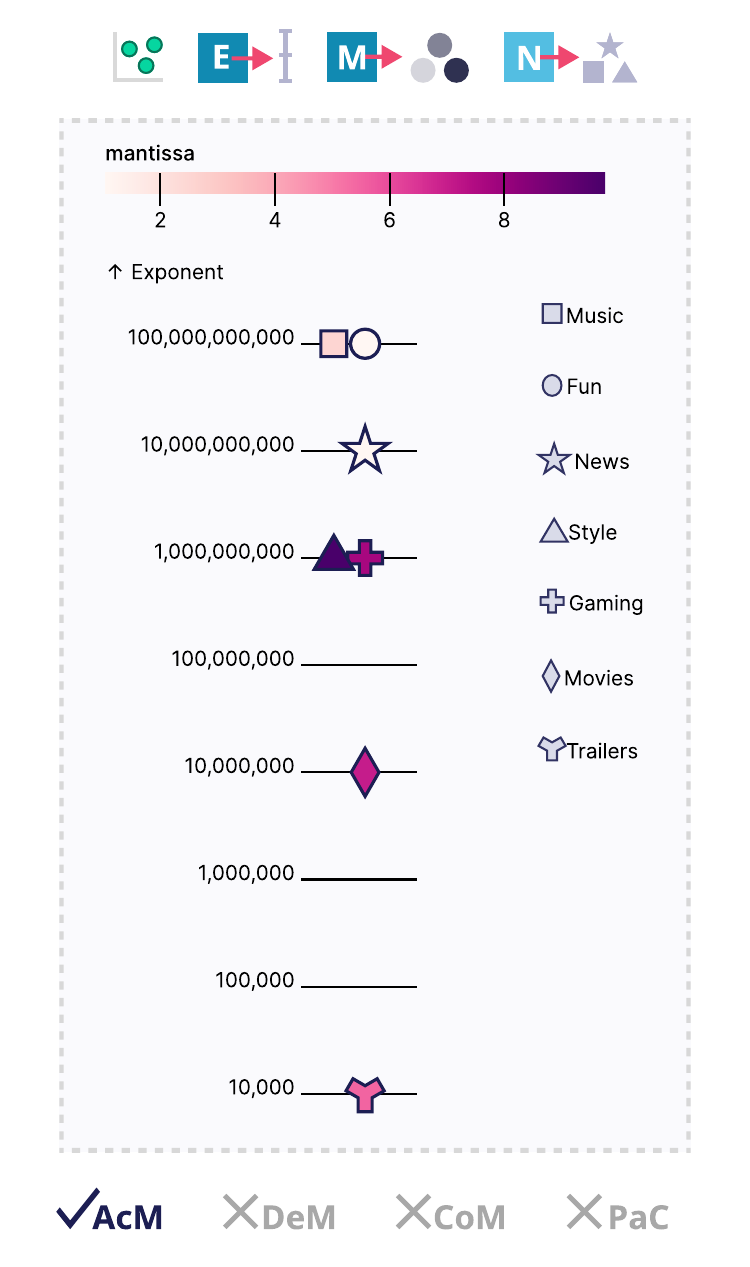}}
        \hspace{2em} % Add some horizontal space between images
        \subcaptionbox{``Impractical for \omvs''
            \label{fig:bad3}}{
            \includegraphics[height=5cm]{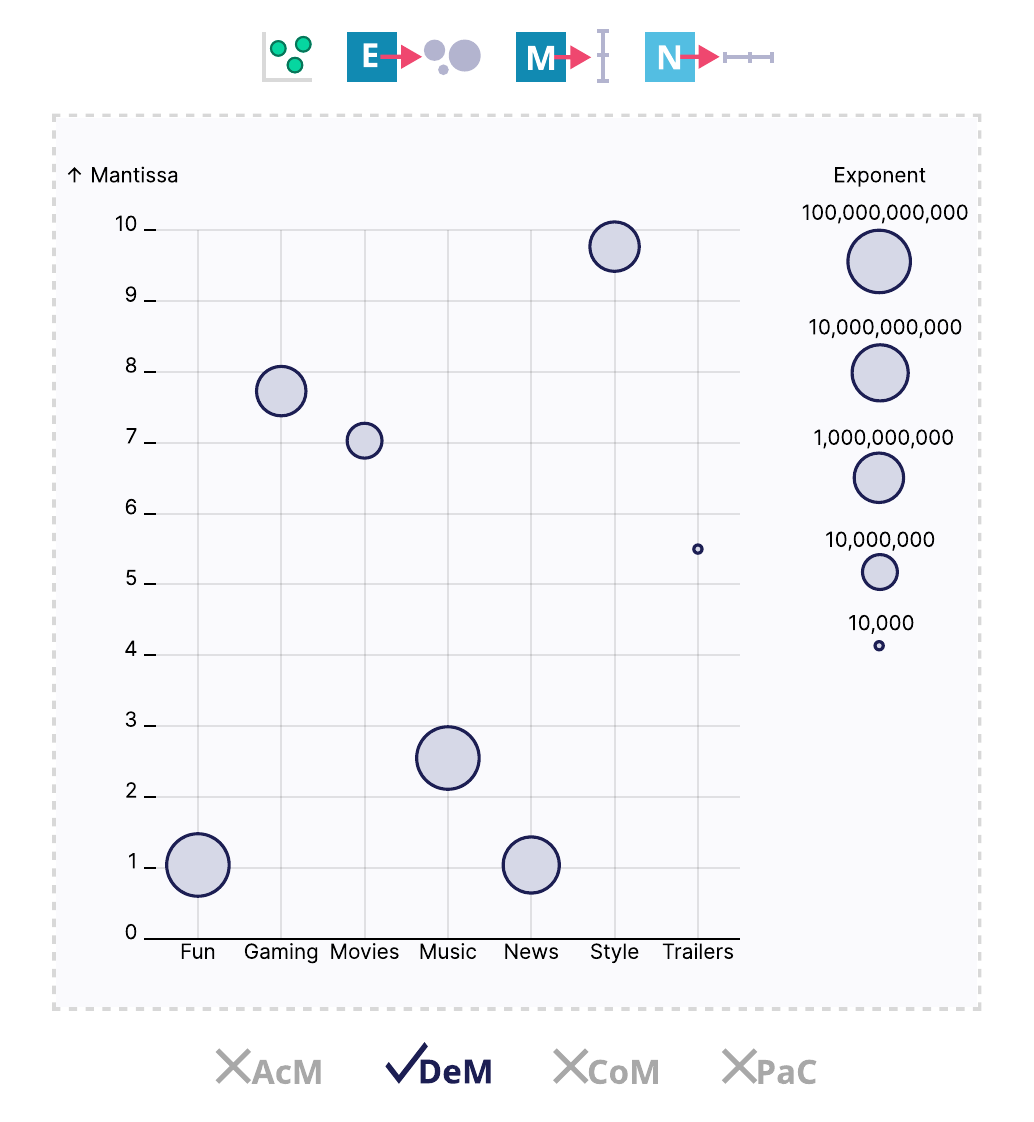}}
        \caption{Examples of combinations that violate guidelines, labeled based on the group to which they belong. \added{At the top, visualizations are described according to the design space (see \autoref{fig:description-of-visualization}), while at the bottom are abbreviations for the guidelines (see \autoref{sec:guidelines}) each visualization follows (\cmark) or violates (\xmark).}}
        \Description{The image shows three visualizations that do not follow the proposed guidelines. The first visualization uses the y-position for the exponent, the x-position for the mantissa, and color hue for a third attribute, violating the guidelines for "Continuity Between Magnitudes" and "Parsimony in Channels." The second visualization uses the x-position for the exponent, color intensity for the mantissa, and shape for the third attribute, adhering only to the guideline of "Accuracy for Magnitude." The third visualization encodes the exponent using area, the mantissa using the y-position, and the categories using the x-position, adhering only to the guideline of "Detail Inside Magnitudes."}
        \label{fig:badAll}
        \vspace{-1em}
\end{figure*}

The qualitative evaluation yielded a set of observed problems and labels for each visualization regarding both encoding and decoding (see \autoref{tab:problems}). Based on these insights, we categorized the 168 visualizations into four groups: 
\begin{description}[nosep]
    \item[\textbf{Effective assignments (25/168):}] Visualizations labeled as suitable for \omvs had no significant encoding issues and only minor decoding caveats that could be resolved with fine-tuning or design enhancements. All visualizations in this category utilized a positional channel (\posy or \row) for the exponent, paired with a highly accurate channel for the mantissa (\posy or \len) aligned in the same direction, ensuring visual continuity. \autoref{fig:goodcomb} shows three examples of these visualizations.
    \item[\textbf{Value retrieval, but no comparisons (7/168):}] Visualizations effective for value estimation but requiring extra effort for quantitative comparisons. While they use highly accurate and discriminable channels for both mantissa and exponent, the lack of continuity forces users to multiply the two to derive the value (\autoref{fig:bad1}).
    \item[\textbf{Magnitude overview, but no detail (55/168):}] Visualizations that effectively communicate the magnitude of a value by using a highly accurate channel for the exponent, but the encoding of the mantissa with a low-discriminability channel (e.g., \colint or \area) requires moderate effort to perform tasks and leads to low confidence in the results (\autoref{fig:bad2}).
    \item[\textbf{Impractical for \omvs (81/168):}] Visualizations that are unsuitable for visualizing \omvs. This group includes those that lack a highly accurate channel for the exponent or exhibit interference between channels, making them unsuitable for tasks including \omvs (\autoref{fig:bad3}). 
\end{description}

% Additionally, it provided valuable insights into potential caveats and trade-offs associated with different visual assignments.

Through our analysis, we identified common patterns that we consider necessary for the effective visualization of \omvs. We present these patterns as design guidelines, which are detailed in the following section. The 168 images used in our evaluation are available in the supplemental material.

\subsection{Design Guidelines}\label{sec:guidelines}
Based on the results of the evaluations, we established four guidelines for effective visualizations. The first, \textbf{Accuracy for Magnitude (AcM)}, requires encoding the exponent with a highly accurate and discriminable channel to avoid significant errors in value estimation and comparisons, which is crucial for visualizing \omvs. The second, \textbf{Detail Inside Magnitudes (DeM)}, stresses the need for the mantissa to also use an accurate and discriminable channel, enabling precise comparisons and estimations of differences and ratios within a single magnitude. Failure to follow this guideline reduces accuracy. \textbf{Continuity Between Magnitudes (CoM)} ensures a smooth transition between exponents by using a mantissa channel that mirrors the numerical relationship between the exponent and mantissa. This ensures a logical progression across values and aids in comparing neighboring exponents. Lastly, \textbf{Parsimony in Channels (PaC)} recommends using the fewest possible channels to represent both attributes. This minimizes channel interference, and allows for the effective integration of additional data attributes into the visualization. 
% We further discuss these guidelines in \autoref{sec:discussion}.

\begin{figure*}[ht]
   \begin{subcaptiongroup}
    \includegraphics[width=0.9\textwidth]{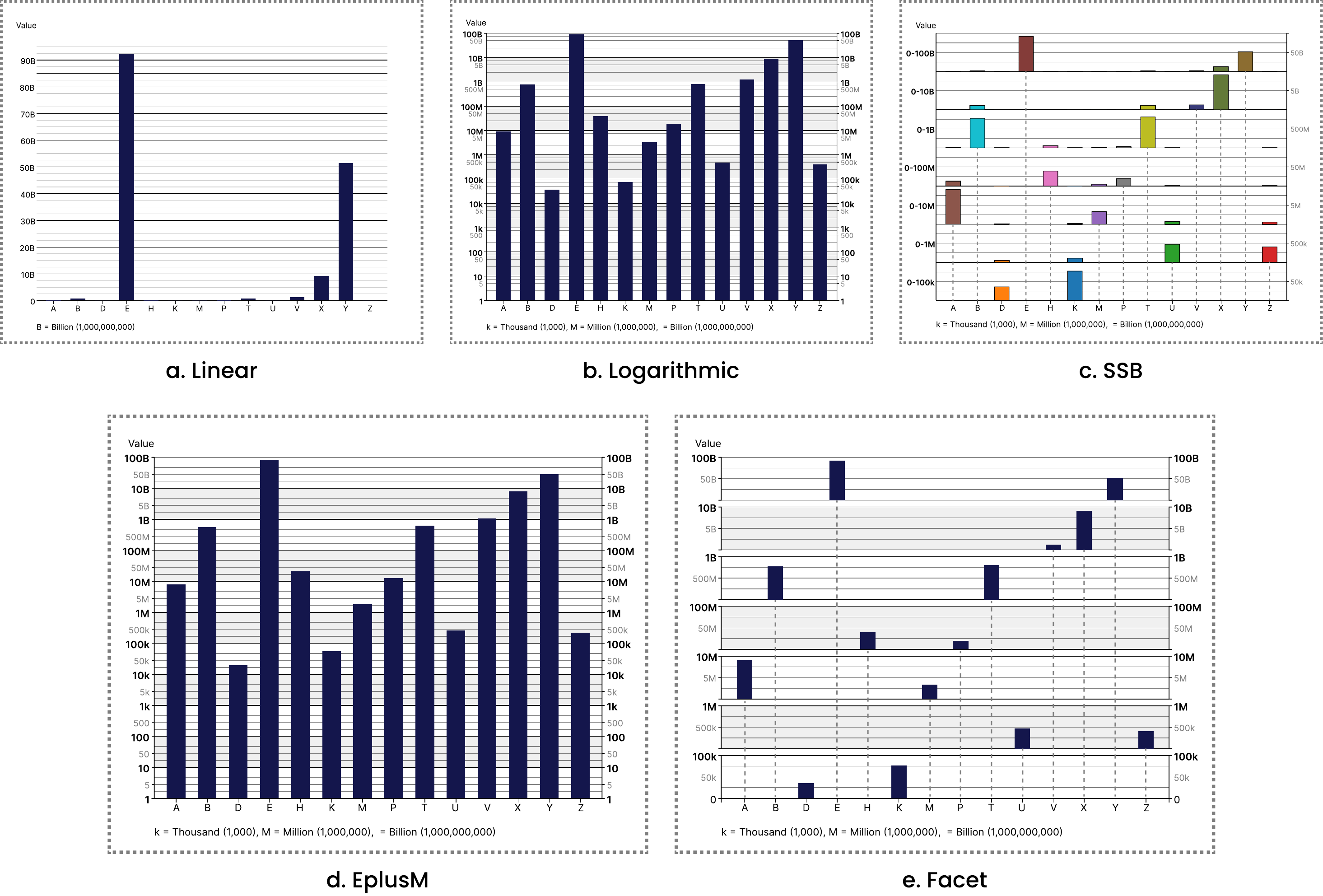}
    \phantomcaption\label{fig:vLinear}
    \phantomcaption\label{fig:vLogarithmic}
    \phantomcaption\label{fig:vSSB}
    \phantomcaption\label{fig:vEplusM}
    \phantomcaption\label{fig:vFacet}
    \end{subcaptiongroup}
    \caption{Visualization designs used for the crowdsourcing evaluation: 
      \subref{fig:vLinear} Linear bar chart,
      \subref{fig:vLogarithmic} Logarithmic bar chart,
      \subref{fig:vSSB} Scale-Stacked Bar Charts (\SSB)~\cite{hlawatsch_scale-stack_2013},
      \subref{fig:vEplusM} EplusM bar chart,
      \subref{fig:vFacet} Facet bar chart. All visualizations display the same dataset.}
    \label{fig:visualizations}
  \vspace{-1.2em}
    \Description{The image shows examples of the visualizations used in our crowdsourcing evaluation. The first is a linear scale bar chart, the second is a logarithmic bar chart, the third is the Scale-Stack bar chart, the fourth is the EplusM scale, and the last is the Facet bar chart. This figure highlights the consistent design choices across the evaluated designs.}
  \end{figure*}

Overall, two encodings for mantissa and exponent that adhere to the mentioned guidelines and appear to result in effective visualizations for \omvs are the \textbf{\EplusM} (see \autoref{fig:vEplusM}) scale, which encodes both the mantissa and exponent with the same positional channel and the \textbf{Facet} (see \autoref{fig:vFacet}), in which the exponent is encoded with row and the mantissa with Y position. To validate that our guidelines can improve the effectiveness of visualizations that separate \omvs into mantissa and exponent, we conducted an empirical study described in the following section.

\section{Controlled Experiment} 
To test whether adhering to our proposed guidelines leads to more effective visualizations, we assessed the effectiveness of EplusM and Facet in a crowdsourcing experiment using Prolific~\cite{prolific}. This study compared \EplusM and Facet with two baseline conditions: linear and logarithmic scales, as well as a well-established visualization from the literature, the \SSB~\cite{hlawatsch_scale-stack_2013}. We selected \SSB because it is the first and most widely adopted visualization to split mantissa and exponent for \omvs, and it fulfills most of our proposed guidelines.

We expect our proposed solutions to outperform \SSB for two primary reasons. While \SSB uses positional channels for both mantissa and exponent, adhering to two of our proposed guidelines, we argue that the repetition of values in each stacked scale violates the guideline of maintaining continuity between magnitudes. Additionally, the use of color to represent categories breaks the guideline of parsimony in the use of visual channels. As \SSB already outperformed Lin and Log~\cite{hlawatsch_scale-stack_2013}, we expect our designs to outperform them as well.

We formulated the following hypotheses:
\begin{itemize}[nosep]
    \item \textbf{H\textsubscript{err}}:\textit{Our designs exhibit similar or lower error rates for tasks on \omv visualizations compared to Lin, Log and \SSB.}
    \item \textbf{H\textsubscript{tim}}:\textit{Our designs exhibit similar or shorter response times for tasks on  \omv visualizations compared to Log and \SSB.} 
    \item \textbf{H\textsubscript{con}}:\textit{Our designs exhibit similar or increased confidence for tasks on \omv visualizations compared to Lin, Log and \SSB.} 
\end{itemize}

\subsection{Data Generation}
In our empirical evaluation, we focused on the example of a national budget, a common case of a dataset with a wide range of values spanning multiple orders of magnitude. Specifically, we used the French National Budget, which includes programs with \omvs ranging from 10 thousand ($e=4$) to 100 billion Euros  ($ e =11$), covering seven orders of magnitude. The use of such data extends the range of orders of magnitudes previously studied (4 for~\cite{hohn_width-scale_2020} and 5 for~\cite{hlawatsch_scale-stack_2013,borgo_order_2014,braun_reclaiming_2023}) and allows the use of real numbers to represent mantissa---unlike previous ones who used only integers.
% Unlike previous studies that have typically assessed visualization designs using narrower value ranges, with exponents ranging from 0 to 5 and integer mantissas~\cite{hohn_width-scale_2020, borgo_order_2014, braun_reclaiming_2023, hlawatsch_scale-stack_2013}, our approach extends this range. 
% By doing so, we can evaluate the performance of different visualization designs across a wider range of \omvs, where the mantissa \( m \) is a real number within the interval \([1, 10)\) and the exponent \( e \) ranges from 4 to 11.

To test the effectiveness of the visualization designs under these conditions, we generated synthetic datasets that reflect this extensive value range. Each dataset included 14 categories, with two categories assigned to each exponent level, similar to previous studies~\cite{hohn_width-scale_2020}. This design allowed for comparisons both within the same exponent and across neighboring exponents. The mantissa for these categories was randomly generated within the range $[1, 10)$, and each category was labeled with a unique letter from the alphabet. In total, we generated 500 statistically equivalent datasets. Every participant performed the trials with a different dataset, to ensure the results are not overly influenced by potential artifacts in the characteristics of one generated dataset per task. While we do not claim ecological validity with these synthetic datasets, they provide controlled conditions that highlight potential challenges and issues related to \omvs in different visualizations.

\subsection{Experiment Stimuli}
Given that our dataset focuses on a national budget, we required a visualization that effectively supports both nominal data (the budgeting categories) and \omvs (the budget amounts for each category). We opted for a line mark, as bar chart designs are among the most common visualization methods for this type of data~\cite{beland}. For the visual channels, we chose the positional encodings of \EplusM and Facet, as these encodings exhibit the least number of problems during our qualitative evaluation and reflect well our proposed guidelines. 
% We deliberately avoided redundant encoding to minimize potential interference effects between visual channels, while ensuring that any observed differences between designs could be attributed to the positional encodings investigated. 
We implemented the visualizations using d3.js~\cite{d3} and ensured consistent size dimensions with Sass~\cite{sass} and React~\cite{react}.

We examined five visualization designs: the Linear Bar Chart, Logarithmic Bar Chart, Stacked-Scale Bar Chart (\SSB), \EplusM Bar Chart, and Facet Bar Chart (\autoref{fig:visualizations}). We refined the \EplusM and Facet Bar Charts using state-of-the-art design guidelines~\cite{munzner2015visualization}, while the \SSB design was faithfully replicated in its original form. To ensure comparability across all visualizations, we standardized key design elements such as the font, font size, grid lines, and overall styling. For instance, we used thick black horizontal lines to separate scales by exponent and thin grey lines to subdivide each scale into four equal parts, corresponding to mantissa values of 2.5, 5, and 7.5, consistent with the \SSB design. In addition, we added a single tick at the mantissa value of 5 in all designs, alongside the existing ticks used to differentiate between exponents. While we maintained a uniform color scheme across all visualization designs, we preserved the original color encoding in the \SSB design. 
% To adhere to the principle of parsimony, we avoided redundant color hue encoding for categories, as the study tasks did not require categorical discriminability. Instead, we achieved discriminability through interface design, 
We believe that the redundant encoding for category in \SSB does not impact user performance, as it is not related to \omvs retrieval or comparison. These consistent design choices ensured a fair assessment of each visualization design's impact on participant performance, avoiding confounding factors.

\begin{figure*}
    \includegraphics[width=0.85\textwidth]{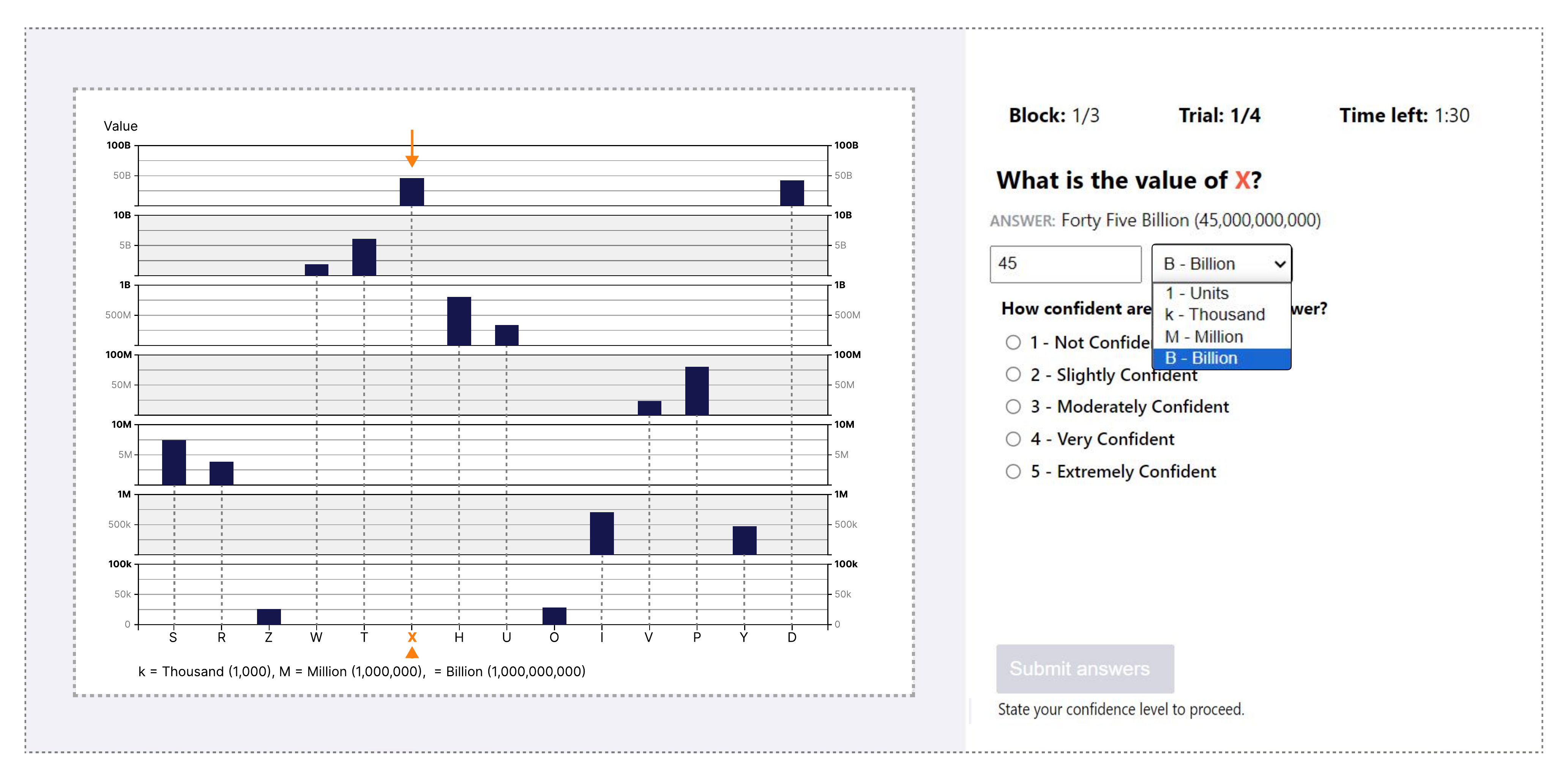}
    \caption{Interface of the tool we developed to conduct the crowdsourcing experiment. Example of Value Retrieval task.}
    \label{fig:interface}
    \Description{The image shows a screenshot from the tool we developed to conduct the crowdsourcing experiment. It illustrates an example of a value retrieval task. On the left, the evaluated visualization is displayed, with orange arrows indicating the selected item for which the participant must retrieve the value. In the top right, information about the current trial, block, and remaining time is shown. Below that is the question the participants must answer, along with input fields for providing their response. Beneath the input fields is a Likert scale with radio buttons to indicate the participant's confidence level in their answer. In the bottom right, there is a button to proceed to the next trial.}
  \vspace{-1.2em}
\end{figure*}

\subsection{Tasks}
We focused on the following low-level perception and cognition tasks for magnitude estimation and quantitative comparison:

\paragraph{Value retrieval} %In the value retrieval task, t
The objective is for participants to estimate a value as accurately as possible. This task forms the first block of our experiment, consisting of four trials. 
% In each trial, participants answer the question: ``What is the value of x?'', where x is a randomly selected item.
\paragraph{Difference estimation} 
%In the difference estimation task, 
Participants aim to estimate the difference between two values as accurately as possible. This task constitutes the second block of our experiment and builds on the approach by Braun et al.~\cite{braun_reclaiming_2023}. We structured the task into three subtasks based on the relationship between the two values: belonging to the \textit{same} order of magnitude, \textit{neighboring} orders of magnitude, or \textit{distant} orders of magnitude. 
% The tasks follow a fixed order of difficulty, with the smallest differences being the easiest to compute. Each subtask includes four trials. During each trial, participants respond to the question: ``What is approximately the absolute difference between x and y?'', where x and y are randomly selected categories that meet the magnitude criteria for the specific subtask.
\paragraph{Ratio estimation} 
%In the ratio estimation task, 
Participants aim to estimate the ratio between two values as accurately as possible. This task constitutes the third block of our experiment, and similarly to the difference estimation task, we structured the task into three subtasks based on the relationship between the two values: \emph{same}, \emph{neighboring}, and \emph{distant}.
%belonging to the same order of magnitude, neighboring orders of magnitude, or distant orders of magnitude. 
% The tasks follow a fixed order of difficulty, with the smallest ratios being the easiest to compute. Each subtask includes four trials. During each trial, participants respond to the question: ``How many times bigger is x in comparison to y ?'', where x and y are randomly selected categories that meet the magnitude criteria for the specific subtask. 

The three blocks were presented in the same order for every participant, from the easiest to the most difficult: 1) Value retrieval, 2) Difference, and 3) Ratio. Within the last two blocks, the three tasks also followed a fixed order from the easiest to the most difficult: 1) \textit{same} order of magnitude, 2) \textit{neighboring} orders of magnitude, and 3) \textit{distant} orders of magnitude. Each of the seven tasks included four trials.

To help participants identify the selected categories, we highlighted them in orange and used arrows positioned above the corresponding bar and below the letter on the x-axis as additional visual cues. The answer input fields align with the format used in the visualization, featuring a numeric input field for values between 1 and 999, along with a dropdown menu to select the appropriate unit (thousands (k), millions (M), or billions (B)). We displayed the participant's response above the input fields in both words and numbers, reinforcing their input. \autoref{fig:interface} shows an example of the interface used for the value retrieval task.

\subsection{Measures}\label{sec:measures}
We evaluated and compared the visualization designs using three measures: accuracy, response time, and subjective confidence. For accuracy, previous studies involving tasks with precisely determinable responses have used two error measurement methods:
% \begin{itemize}
    \paragraph{Absolute Relative Error = $\left|1 - \frac{\text{response}}{\text{correct}}\right|$} This error metric, used by Hlawatsch et al.~\cite{hlawatsch_scale-stack_2013} and Braun et al.~\cite{braun_reclaiming_2023}, measures the relative deviation from the correct value. It has some advantages for measuring accuracy with \omvs as it takes into account the characteristics of logarithmic scaled data. For example, $\left|1 - \frac{10}{100}\right|$ and $\left|1 - \frac{1000}{10000}\right|$ both yield an error of 0.9, despite the absolute errors (90 and 9000) being different. However, this method is asymmetrical. It produces a lower error when the participant underestimates the correct value, and a higher error when they overestimate. For example, if the correct value is 100 and the response is 10, the error is 0.9, but if the correct value is 10 and the response is 100, the error jumps to 9.
    
    \paragraph{Log Relative Error = $\log_{10}\left(\frac{\text{response}}{\text{correct}}\right)$} This error metric (hereafter abbreviated in \LRE), used by Höhn et al. ~\cite{hohn_width-scale_2020}, calculates the logarithmic relative difference between the correct and response values. It is symmetric ~\cite{Törnqvist_1985}, meaning that the same order-of-magnitude error yields the same value regardless of whether the participant overestimates or underestimates. Tofallis~\cite{tofallis_better_2015} further supports that it is the most accurate measure of relative differences for heteroscedastic data, which is particularly relevant for \omvs, where the response and the correct value can differ significantly in orders of magnitude.
% \end{itemize} 

Given these considerations, we used \LRE to measure accuracy, comparing participants' responses to the ``visually correct value,'' defined as the highest level of detail the visualization conveys. Based on our data range and visualization dimensions, the visually correct value was rounded to two digits after mantissa. For analysis, we took the absolute Log Relative Error since the direction of the error was not of interest. The \LRE is 0 when the response matches the correct value, and 1 when the error is off by one exponent. In value retrieval tasks, a Log Relative Error between 0 and 1 indicates an error in estimating the mantissa.
% where a 1-digit mistake in the mantissa results in an error ranging between 0.05 and 0.3.
Since the logarithm of 0 is undefined, we replaced any participant responses of 0 (equivalent to NaN or null) with 1 for calculation purposes. 

% In order to better interpret the errors, an exponent error of 1 indicates a log relative error greater than 1 and a mantissa error of 1 corresponds to a logarithmic error between 0.05 and 0.3.
% while a mantissa error of 0.5 translates to a logarithmic error within the range of 0.02 to 0.16.

\subsection{Experimental Design}
We conducted the study online using the Prolific.com~\cite{prolific} service. We employed a between-subjects design for the different visualizations and a within-subjects design for the different tasks~\cite{CHARNESS20121}. This approach reduced the number of trials per participant, allowing us to distribute tasks across various data conditions and enabling a more in-depth analysis of quantitative comparisons for values spanning multiple orders of magnitude. To conduct the experiment, we developed a custom tool. 
% using React~\cite{react} for the front end and PHP~\cite{PHP} for the server-side operations.

\subsubsection{Pilot}
We conducted two pilot studies before the main experiment. In the first pilot, we recruited 30 participants through personal contacts to gather feedback and estimate the approximate duration of the experiment. Based on their input, we made several adjustments to both the instructions and the interface design. For the second pilot, we involved 22 users from Prolific to identify any technical issues that needed to be resolved before the main study. This pilot revealed responsiveness issues with specific screen dimensions, which we promptly addressed. 

\subsubsection{Participants and exclusion criteria}
We conducted a power analysis to determine the number of participants required for our study. Using G*Power~\cite{GStarPower}, we assumed an effect size of 0.25, an alpha level ($\alpha$) of 0.05, and a power of 0.81. The analysis was based on 5 groups and 4 measurements per task (trials), with a correlation of 0.5 between measurements. Based on these parameters, the total sample size needed is 130, which equates to 26 participants per visualization. We recruited participants in an IRB-approved study on Prolific.co who are between the ages of 18 to 70, have a normal or corrected-to-normal vision, speak fluent English, had an approval rate 99\%, and had completed more than 10 previous studies. We excluded participants who failed attention check questions, did not complete the study, had three consecutive missing answers, or achieved an average performance rating of fewer than two stars (indicating a lack of understanding of the experiment). We also excluded one participant, who wrote in the feedback section that he encountered technical problems during the experiment. In total, data were collected from 150 participants. After excluding 10 participants for failing attention checks, 3 for missing trials, 4 for low performance, and 1 who faced technical issues, 132 participants remained for analysis. Successful participants received compensation at an average rate of £9.97 per hour, with a median completion time of 31 minutes. The participants' mean age was 25 years, with 41 women, 55 men, and 36 who did not consent to share this information. 
%They came from 21 different countries, with the majority (37\%) being from Europe.

\subsubsection{Crowdsourcing Quality Control}
To ensure the integrity of our online experiment, we established strict hardware and software requirements for participants, aiming to reduce as much as possible the influence of confounding variables. Specifically, we required devices with a screen resolution of at least 1440 pixels in width and 850 pixels in height. We calculated visualization sizes in centimeters, accounting for each device's pixel ratio and zoom level. To maintain consistency in the visual presentation, we restricted participants from adjusting their browser's zoom level. If the zoom level deviated from 100\%, a full-screen message prompted them to reset it before continuing the experiment. We included a comprehension test after the instructions, allowing participants two attempts to answer correctly, and embedded three attention checks throughout the experiment. Additionally, we tracked how many times participants changed or minimized their window as an extra measure of attention. Finally, we incorporated a star rating for performance as a gamification element to enhance engagement, and we awarded a bonus payment to the top 5\% of performers in each condition. Potential uncontrollable external influences such as distracting environmental noise was averaged out due to the large number of participants.

\subsubsection{Procedure}

The entire study lasted approximately 30 minutes and was decomposed into 5 phases:
\paragraph{Introduction and consent} Each participant began by reading an overview of the study's goals, requirements, and constraints. They then read the consent form and decided whether to participate in the experiment.
% The study required them to perform the tasks in full-screen mode without navigating back or changing the browser's zoom level. The instructions emphasized staying focused, responding quickly and accurately, and avoiding the use of a calculator. The study offered a bonus payment to top performers. Finally, participants read the consent form and decided whether to participate in the experiment.
\paragraph{Subjective Numeracy Test} Each participant assessed their own ability with various mathematical tasks using the Short Subjective Numeracy Scale~\cite{McNaughton2011-gu, McNaughton2015-or}. Researchers demonstrated this scale is a highly accurate proxy to measure numeracy. We added two additional Likert scale questions regarding familiarity with logarithmic scales, as well as an attention check~\cite{crowdsource}.
\paragraph{Instructions} Each participant saw detailed instructions about how to read the specific visualization design assigned to them and what they had to do during the experiment. At the end of the instructions, there were two comprehension check questions to ensure the understanding of the instructions. They had two chances to answer correctly; if they failed, the study was stopped. % and based on Prolific policies they were requested to return their submission. 
The median time spent in this section was 8.28 minutes.
\paragraph{Trials} 
% Each participant completed three blocks of trials, each for a different task. 
%The first block involved value retrieval with four trials, the second involved difference estimation with 12 trials (four per subtask), and the third involved ratio estimation with 12 trials (four per subtask). 
Participants had one and a half minutes to answer each question and indicate their confidence level for their answer. After this time, the system automatically advanced to the next trial. To prevent participants from rushing through the trials, a five-second minimum viewing time was enforced before they could proceed, ensuring they had sufficient time to engage with the visualizations. After submitting their answers, participants received a performance rating in the form of stars, offering feedback on their accuracy and increasing engagement.
%The tasks were organized in a fixed order for all participants, arranged by increasing difficulty. 
Each participant received a unique dataset, with different visualized data for each trial. At the end of each block, participants saw a screen displaying a summary of their performance using the same stars system and introducing the next task. These screens also included attention check questions. The median time spent on this section was 16.42 minutes.
\paragraph{Feedback} At the end of the experiment, participants were asked to rate their overall experience using a 5-point Likert scale, ranging from ``Very Negative'' to ``Extremely Positive.'' Additionally, they were invited to provide written feedback on the effectiveness of the visualizations for the various tasks, any problems they encountered, and suggestions for improving the visualization designs.

The tool used for the evaluation and the instructions for each visualization are included in the supplemental material.
% \begin{figure*}[ht]
%     \raggedleft
%     \includegraphics[width=0.9\textwidth]{img/log-error.pdf}
%     \caption{Log \added{Relative} Error \added{(\LRE)} per visualization design and task, with error bars showing the 95\% CIs.}
%     \label{fig:error-CI}
%     \Description{The image shows the geometric mean for the log relative error, showing 95\% CIs, per visualization design and task. Overall the lin barchart has the higher error rates across all the tasks. Facet and EplusM have the least mean log relative error per task, with better performance for estimating ratio betwenn same and neighboring magnitudes. \SSB and log appear to have similar performance across tasks.}
%     \vspace{-1.2em}
% \end{figure*}

% \begin{figure*}[ht]
%     \raggedleft
% \includegraphics[width=0.9\textwidth]{img/error-dif-ci.pdf}
%     \caption{\changed{Pairwise differences in Log Relative Error (\LRE) between Log, \SSB, \EplusM, and Facet per task with 95\% CIs}.}
%     \label{fig:error-dif-CI}
%     \Description{The image compares the log relative error between different pairs of the evaluated visualizations. \EplusM and Facet perform similarly across tasks. Facet appears to perform better than \SSB and the logarithmic scale for estimating differences and ratios between values in the same or neighboring magnitudes. For \EplusM, the observed effect over \SSB and the logarithmic scale is larger in the ratio estimation task. We did not observe a significant effect between Log and \SSB.}
%   \vspace{-1.2em}
% \end{figure*}

\begin{figure*}[ht]
    \includegraphics[width=0.9\textwidth]{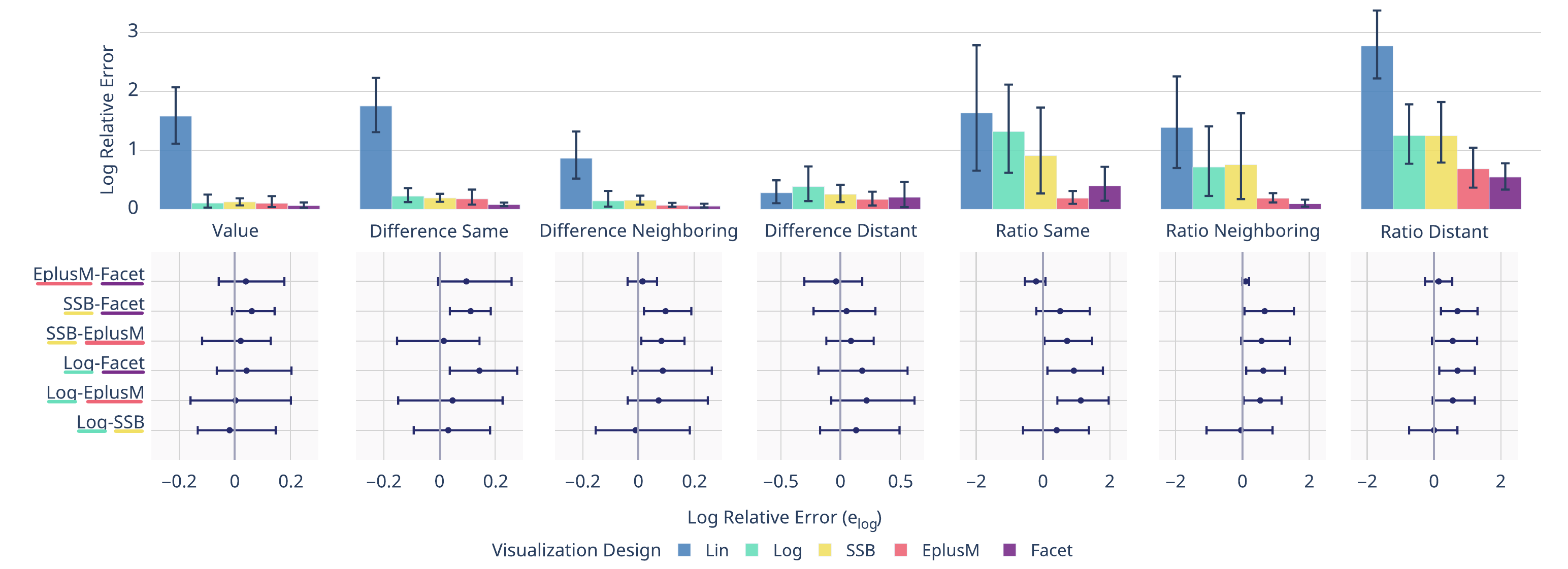}
    \caption{The bar chart shows the Log \added{Relative} Error (\LRE) per visualization design, grouped by task, with error bars representing the 95\% CIs. For each task, we display the pairwise differences in \LRE between Log, \SSB, \EplusM, and Facet with 95\% CIs.}
    \label{fig:error-CI}
    \Description{The image combines two charts, based on our error analysis. The one at the top, is a grouped bar chart that shows the geometric mean for the log relative error, showing 95\% CIs, per visualization design and task. Overall the lin barchart has the higher error rates across all the tasks. Facet and EplusM have the least mean log relative error per task, with better performance for estimating ratio betwenn same and neighboring magnitudes. \SSB and log appear to have similar performance across tasks. The one at the bottom, shows for each task of the grouped barchart the pairwise differences between the evaluated visualizations. \EplusM and Facet perform similarly across tasks. Facet appears to perform better than \SSB and the logarithmic scale for estimating differences and ratios between values in the same or neighboring magnitudes. For \EplusM, the observed effect over \SSB and the logarithmic scale is larger in the ratio estimation task. We did not observe a significant effect between Log and \SSB.}
    \vspace{-1.2em}
\end{figure*}

\begin{figure*}[ht]
\includegraphics[width=0.9\textwidth]{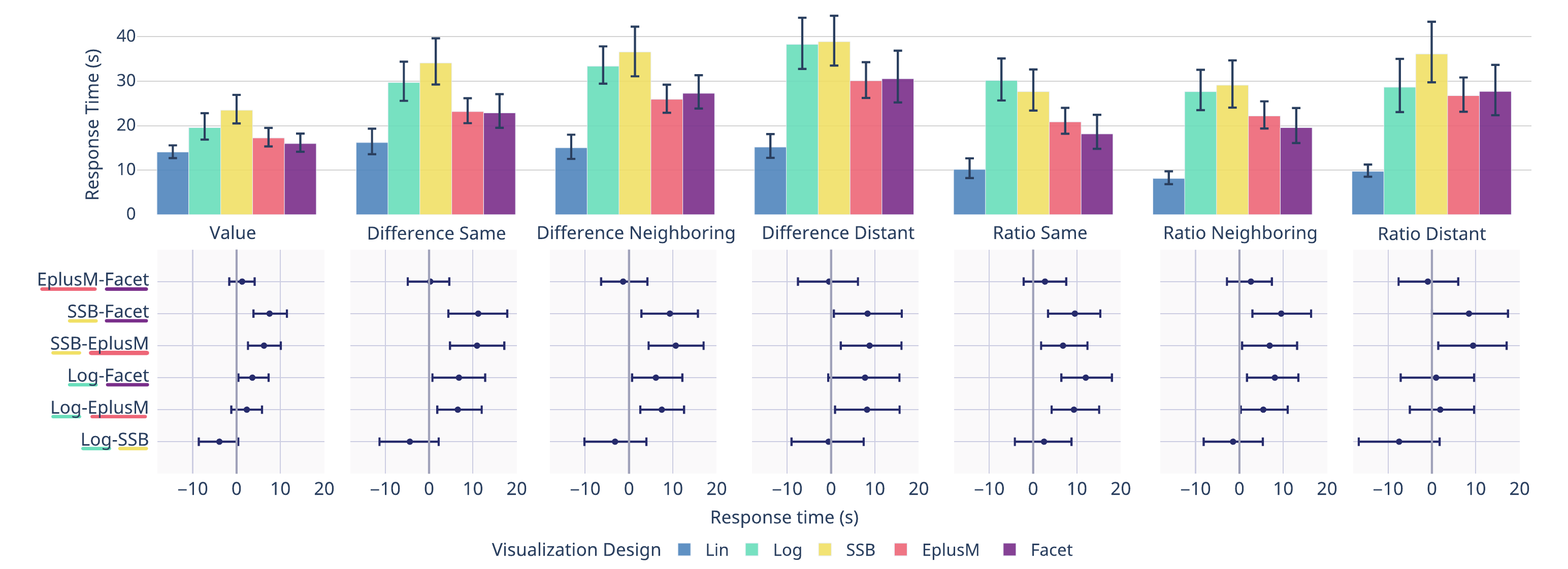}
    \caption{\changed{The bar chart shows the Response Time in seconds per visualization design, grouped by task, with error bars representing 95\% CIs. For each task, we display the pairwise differences between Log, SSB, \EplusM, and Facet with 95\% CIs.}}
    \label{fig:time-CI}
    \Description{The image combines two charts, based on our response time analysis. The one at the top, is a grouped bar chart that shows the geometric mean of the response time, with 95\% confidence intervals, for each visualization design and task. Overall,  has the shortest response time across all tasks. Facet and EplusM follow, with better performance in estimating ratios between values in the same or neighboring magnitudes. Log and \SSB have similar performance across tasks. The one at the bottom, shows for each task of the grouped barchart the pairwise differences between the evaluated visualizations.\EplusM and Facet perform similarly across tasks. Facet and \EplusM, consistently outperformed \SSB and Log across the different tasks, with larger effect observed for ratio estimation between values in the same magnitude. We did not observe a significant effect between Log and \SSB.}
  \vspace{-1.2em}
\end{figure*}

\subsection{Results}
 The final number of participants included in our study after exclusion criteria was 26 for Lin, 27 for Log, 26 for \SSB, 26 for \EplusM, and 27 for Facet. Since all included participants completed every task, our analysis is based on 132 observations, corresponding to the total number of participants. There is a weak correlation (Spearman Correlation: -0.19, p<0.05) or no correlation between the experiment measurements and participants' subjective numeracy or familiarity with logarithmic scales. Therefore, we do not consider these aspects in our analysis. In the following paragraphs, we present the results of our experiment per measurement. We follow current APA recommendations~\cite{APA2020Manual} and report our results using estimation techniques with effect sizes and confidence intervals (CIs), and thus avoid words such as ``significant'' in the discussion of our results~\cite{dragicevic_statistics}. \added{We used bootstrapping to compute all the CIs. } The anonymized experimental data, the analyses from this experiment, and the tables with the results are available in the supplemental material.

\subsubsection{Accuracy}

\changed{For accuracy, we used the \LRE measurement described in \autoref{sec:measures}. For each participant, we aggregated their four trials using the median error as it offers a robust measure of central tendency, reducing the influence of extreme values caused by large errors. 
% This approach provides a reliable indicator of participants’ performance. 
Following Tofallis~\cite{tofallis_better_2015}, we analyzed the \LRE for each visualization and task using the geometric mean.   The results (cf. \autoref{fig:error-CI}) indicate that Lin is the least accurate for most tasks, except for estimating differences between distant magnitudes. Consequently, and to maintain the readability of the charts, we chose not to perform pairwise comparisons for Lin. }

% Our results show that the Facet and \EplusM exhibit lower error rates (H\textsubscript{err}), particularly for tasks requiring quantitative comparisons.
\secondChanged{\textbf{Our results suggest that accuracy varies across tasks, with stronger effects observed for quantitative comparisons, where Facet and \EplusM seem to exhibit lower error rates (H\textsubscript{err}).}} We present a summary of key observations per task below:
\begin{itemize}
    \item \textbf{Value retrieval:}  Accuracy appears similar across Log, \SSB, \EplusM, and Facet.
    \item \textbf{Difference estimation:} For differences within the same magnitude, Facet demonstrates the highest accuracy, outperforming both \SSB and Log. Similarly, for differences between neighboring magnitudes, Facet and \EplusM outperform \SSB and Log, with moderate evidence supporting this trend. However, for differences between distant magnitudes, the results are largely inconclusive.
    \item \textbf{Ratio estimation:} Evidence suggests that \EplusM and Facet outperform Log and \SSB across all magnitude conditions, with greater effects observed between \EplusM and Log in the same magnitude condition.
    % , showing the limitations of logarithmic scale to support effectively ratio tasks between values that belong in the same magnitude. 
\end{itemize}

\subsubsection{Response Time}

\changed{For each participant, we computed the geometric mean response time across their four trials. We selected the geometric mean because it effectively handles the skewed data typically seen in response times, reducing the influence of extreme values. \autoref{fig:time-CI} presents the results. The results show that Lin has the fastest performance across all tasks.  However, due to its consistently low accuracy, we chose not to perform pairwise comparisons for Lin.}

\secondChanged{\textbf{Our results indicate that Facet and EplusM appear to be faster than \SSB and Log (H\textsubscript{tim}) across tasks, with a stronger effect observed for quantitative comparisons between values within the same or neighboring magnitudes.}} When comparing SSB and Log, the direction of the effect is uncertain but likely favors logarithmic scales, except in the case of ratio estimation for values within the same magnitude.

% \EplusM and Facet show very similar response times across tasks, with both consistently appearing faster than \SSB.
% When comparing Facet and \EplusM with Log, both seem to perform better in most tasks.
% We notice a strong effect when estimating ratios between values within the same or neighboring magnitudes and a moderate effect for value retrieval and difference estimation.
% However, we were not able to observe an effect for ratio estimation of values in distant magnitudes. 
% When comparing SSB and Log, the direction of the effect is uncertain but likely favors logarithmic scales, except in the case of ratio estimation for values within the same magnitude.
%with a smaller effect for value retrieval and difference estimation but a more pronounced effect for estimating ratios between values within the same magnitude (Facet: geometric mean difference of 12 seconds, 95\% CI [5.6, 17.7]; \EplusM: geometric mean difference of 9.3 seconds, 95\% CI [4, 15.3]) and neighboring magnitudes. 

\begin{figure*}[ht]
\includegraphics[width=0.9\textwidth]{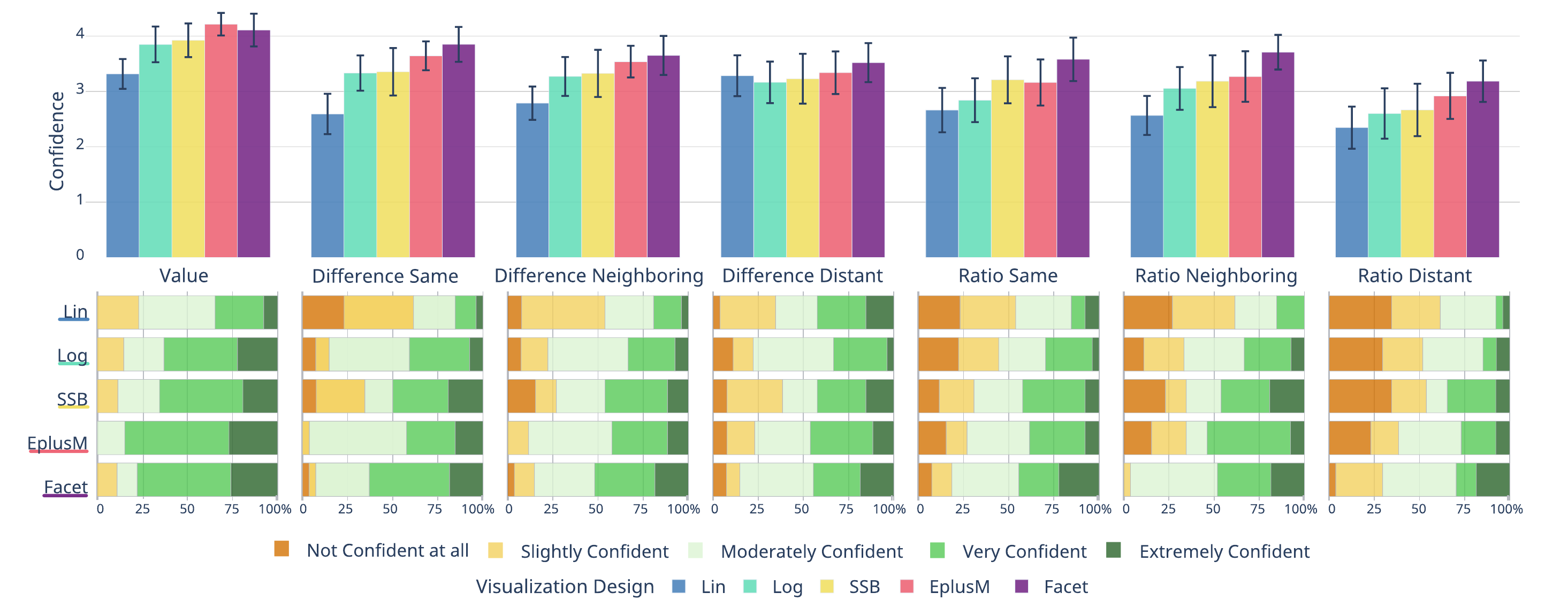}
    \caption{The bar chart on top shows the Confidence per visualization design, grouped by task, with error bars showing 95\% CIs. For each task, the stacked bar charts on the bottom show the distribution of confidence for all trials, per design.}
    \label{fig:confidence}
  \vspace{-1.2em}
   \Description{The image combines two charts, based on our confidence analysis. On the top, the grouped bar chart shows the mean subjective confidence, with 95\% confidence intervals, for each visualization design, grouped by task. Overall, the linear bar chart has the lowest confidence rates across all tasks. Facet and EplusM consistently have the highest mean confidence across tasks. \SSB and the logarithmic chart show similar confidence levels across tasks, with some exceptions where \SSB slightly outperforms. On the bottom, the stacked bar charts  provide an overview of the Likert-scale answer distributions for confidence across all trials, tasks, and designs. Overall, confidence seems to vary based on the tasks, decreasing as task difficulty increases. The linear bar chart consistently received the majority of "Not Confident at All" and "Slightly Confident" ratings across tasks. The majority of "Very Confident" and "Extremely Confident" responses were associated with Facet.}
\end{figure*}

% \begin{figure*}[ht]
% \raggedleft
% \includegraphics[width=0.9\textwidth]{img/confidence.pdf}
%     \caption{Distribution of confidence responses for all trials, grouped by visualization and task.}
%     \label{fig:confidence}
%   \vspace{-1.2em}
%    \Description{The image provides an overview of the Likert-scale answer distributions for confidence across all trials, tasks, and designs. Overall, confidence seems to vary based on the tasks, decreasing as task difficulty increases. The linear bar chart consistently received the majority of "Not Confident at All" and "Slightly Confident" ratings across tasks. The majority of "Very Confident" and "Extremely Confident" responses were associated with Facet.}
% \end{figure*}

% \begin{figure*}[h]
% \raggedleft
% \includegraphics[width=0.9\textwidth]{img/confidence-ci.pdf}
%     \caption{Confidence per visualization design and task, with error bars showing 95\% CIs.}
%     \label{fig:confidence_ci}
%   \vspace{-1.2em}
%    \Description{The image shows the mean subjective confidence, with 95\% confidence intervals, for each visualization design and task. Overall, the linear bar chart has the lowest confidence rates across all tasks. Facet and EplusM consistently have the highest mean confidence across tasks. \SSB and the logarithmic chart show similar confidence levels across tasks, with some exceptions where \SSB slightly outperforms.}
% \end{figure*}

\subsubsection{Confidence}

%\autoref{fig:confidence} provides an overview of the Likert-scale answer distributions for confidence across all trials, tasks, and designs.
\changed{For the visualization and analysis of Likert-scale responses, we follow the recommendations of South et al.~\cite{SouthSVDB22}.  \autoref{fig:confidence} (bottom) provides an overview of the Likert-scale answer distributions for confidence across all trials, tasks, and designs. In analyzing the results, we treat the responses as interval data, assuming equal distances between points. This intervalist approach is consistent with the analysis used for objective measures. We aggregate confidence ratings per participant by calculating the mean confidence level for each task. We then compute the overall mean and 95\% confidence intervals for all visualizations per task, as shown in \autoref{fig:confidence} (top). 
% The categorical responses were rated on a scale from 1 to 5, where 1 represents ``Not confident at all'' and 5 indicates ``Extremely confident.'' A higher score reflects greater confidence. 
This confidence rating serves as an indicator of the subjective effectiveness of the visualizations in supporting task performance, complementing the objective measures.}

\secondChanged{\textbf{Our results suggest that using Facet and \EplusM may lead to increased user confidence across tasks (H\textsubscript{con}).}} Log and \SSB perform similarly across tasks with exception the estimating ratios within the same magnitude, where \SSB appears to have slightly increased confidence. Log and SSB perform similarly, except for estimating ratios within the same magnitude, where \SSB shows slightly higher confidence. Lin demonstrates the lowest confidence across tasks, except for estimating differences between distant magnitudes, where it seems to outperform both Log and \SSB.

\subsection{Result Discussion}\label{sec:discussion}
Overall, the result of the experiment demonstrates that both Facet and \EplusM are effective for visualizing \omvs, offering comparable or lower error rates (\textbf{H\textsubscript{err}}), similar or faster response times (\textbf{H\textsubscript{tim}}), and equal or higher confidence levels (\textbf{H\textsubscript{con}}), compared with Lin, Log, and \SSB. These findings suggest that the visualizations identified as ``Effective assignments'' in our qualitative evaluation, which adhere to the proposed design guidelines, not only match but may surpass the performance of log scale bar charts and state-of-the-art solutions like \SSB across key effectiveness metrics. This advantage is particularly pronounced for quantitative comparisons involving values of similar and neighboring magnitudes.

Our results align with previous studies that used the same error measurement~\cite{hohn_width-scale_2020}, where linear, logarithmic, and SSB charts showed similar performance in terms of time and accuracy. However, they differ from experiments that employed different error measures~\cite{hlawatsch_scale-stack_2013}, where \SSB significantly outperformed Log for ratio estimation tasks. This discrepancy may be attributed to the use of Log Relative Error, which, while advantageous over Absolute Relative Error in some contexts~\cite{tofallis_better_2015}, appears to favor logarithmic scale evaluation. For instance, a 1-digit mantissa error translates to 0.3 between values of 1 and 2 but drops to just 0.05 (six times smaller) between 8 and 9. This suggests that Log Relative Error benefits logarithmic scales, as errors at higher values--- where ticks are denser and retrieving values is more difficult---are weighted less than equivalent errors at lower values within the same interval between two exponents. However, to our knowledge, there is no proposed relative error measure in the literature that addresses this bias and is suitable for analyzing data with high heteroscedasticity. Despite this, we believe the choice of error measure did not affect the comparison between our proposed designs and \SSB, as all Facet, \EplusM, and \SSB use linear scales between the orders of magnitude.

Furthermore, our results demonstrate that it is essential to evaluate the effectiveness of visualizations per task, as both the task and the magnitude difference between compared values influence performance. For instance, in value retrieval tasks, we didn't observe important difference in accuracy between the evaluated designs. However, the effect size was more pronounced in ratio estimation tasks, particularly for values within the same or neighboring magnitudes. Participants reported difficulties with logarithmic scales in these tasks, as reflected in free-text feedback, where one participant noted, \textit{``Since the bars don't show the actual value in a more ordinate way, it was hard for me to calculate the ratio and be more precise in some cases where the jump was bigger.''} Interestingly, although linear bar charts did not perform well in most tasks, participants were able to estimate differences between values in distant magnitudes. This can be attributed to the high probability of having at least one bar within the visible range of linear scales (three larger exponents) in this task. This visibility allowed participants to estimate the difference by focusing on the value of the visible bar, as the difference between distant values was considered negligible.

\section{General Discussion, Limitations, and Future Work}

\added{The objective of our work is to effectively visualize \omvs in overview, inspired by the natural human tendency to conceptualize large numbers as a combination of two components, such as mantissa and exponent. Unlike many design space explorations in visualization literature~\cite{javed2012, brehmer2017, Schulz2011} that build on existing designs, we adopted a different systematic approach, which we believe can be generalized to other scenarios. Our exploration was grounded in the principles of the Grammar of Graphics, which provided a structured foundation for specifying the dimensions of our design space and facilitated the systematic generation of a wide range of visualizations. We then incorporated insights from numerical and graphical perception studies to derive design requirements and constraints, refining the space to a set of viable solutions. These solutions were generated and evaluated using an inspection-based qualitative method to identify potential encoding and decoding challenges, ultimately leading to the development of a set of design guidelines.} 

\changed{Among the solutions, collinear positional encodings for both mantissa and exponent (\EplusM and Facet) emerged as the most effective. They provided highly accurate and discriminable channels for both the exponent (Accuracy for Magnitudes) and mantissa (Detail inside Magnitudes), preserved the natural continuity between them (Continuity Between Magnitudes), and supported the integration of additional data attributes while minimizing interference effects (Parsimony in Channels). A crowdsourcing evaluation demonstrated that \EplusM and Facet performed as well as, or better than, existing solutions for specific tasks, validating our proposed guidelines and reinforcing the importance of separating \omvs into mantissa and exponent for accurate quantitative comparisons. These contributions advance the visualization literature on \omvs and enhance the state of the art.}

\added{Previous studies have already proposed two collinear positional designs, \SSB and \OML, which align closely with our design guidelines and bear similarities to the Facet and \EplusM visualizations that emerged from our exploration. Both \SSB and Facet encode the exponent using rows; however, they differ in how they handle the mantissa. \SSB resets the range within each row starting from 0, whereas Facet encodes the mantissa in \posy within the range $[1,10)$. While \SSB’s approach supports across-magnitude comparisons by enabling users to interpret rows independently, it disrupts continuity between magnitudes and  increases visual clutter—factors likely contributing to its higher response times in most tasks. The positional scales of the \OML and \EplusM designs differ primarily in how they visually represent the mantissa and exponent along \posy. \OML separates these components, requiring users to multiply them, while \EplusM integrates both into a single value. This subtle difference could facilitate users' interpretation, as they do not need to understand the concept of mantissa and exponent or perform mental calculations to determine the final value. Considering the visual similarities with prior work, we do not claim novelty for the \EplusM and Facet designs, but regard them as refined versions and effective representations of our design guidelines.} 

While our study addressed key questions regarding \omv visualizations, it also raised new ones and was unable to resolve all challenges. We did not attempt to explore the full design space of possible \omv visualization designs, and thus our proposed design space is consistent but not complete. By focusing on a simplified subset---tabular data, Cartesian layouts, single-mark visualizations, and no redundant encoding---we were able to systematically investigate perceptual and cognitive challenges related to separating \omvs into mantissa and exponent. This choice served as an essential first step in understanding the visual interplay between these two components before expanding into a broader design space. Although our guidelines may not represent a complete set of rules for effective \omv visualizations, we believe they offer a useful framework for designers and researchers to explain their design rationale and identify potential encoding and decoding issues.

Future research should build on these findings by exploring a wider variety of data types, incorporating redundant encoding or interactivity, and assessing additional tasks. Future studies should also expand evaluations to more complex tasks, such as identifying correlations or supporting decision-making. Additionally, this study focused primarily on the perceptual and cognitive aspects of visualization, leaving out higher-level considerations such as memorability, trust, aesthetics, and user engagement. \secondChanged{Future research should evaluate the real-world applicability and differences between \EplusM and Facet, particularly in their effectiveness at communicating \omvs to a general audience without prior training.}  We believe that our study results generalize to other domains, and mantissa-exponent visualizations could prove beneficial for networks, fields, maps, and specialized visualization applications.

We hope our work inspires visualization designers and researchers to develop novel and effective \omv visualization designs, following and extending our proposed guidelines. We also encourage system developers to implement operators that can easily separate \omvs into mantissa and exponent components. We envision support for \omvs in popular visualization systems, requiring minor adjustments: (1) introducing the \EplusM scale, (2) providing functions to extract exponent and mantissa parts of a value 
% (e.g., through functions $E(v), M(v)$ or syntax such as $v.e, v.m$)
, and (3) introducing a new quantitative type (e.g., \texttt{omv}) in addition to \texttt{quantitative}, allowing for automatic use of the \EplusM scale when appropriate and enabling syntax for accessing the individual parts.

\section{Conclusion}
This work explores the design of mantissa-exponent visualizations for \omvs. We proposed a design space focused on tabular data, informed and constrained by graphical perception principles. To better understand the interplay between mantissa and exponent, we qualitatively evaluated candidate visualizations within this space, identifying encoding and decoding challenges that could impact their effectiveness. Based on these evaluations, we derived design guidelines and validated them through a quantitative comparison of two visualizations aligned with these guidelines against traditional linear and logarithmic bar charts, as well as a state-of-the-art visualization. Our results demonstrate that adhering to our proposed guidelines and separating \omvs into mantissa and exponent components improves performance in quantitative comparisons. Floating-point numbers made possible calculations with \omvs on computers; scientific notation allows human reading of \omvs; our work contributes to generalizing the use of mantissa-exponent visualizations for \omvs.

\begin{acks}
The authors wish to thank 
Ludovic David for his support in the development and bulk generation of images,
Olivier Gladin for his support with the wall-size display, 
Lorenzo Ciccione,
and Emanuele Santos for their useful feedback and thoughtful comments. 
This work has been partially supported by Berger-Levrault.
\end{acks}

\section*{Supplemental Materials}
\label{sec:supplemental_materials}
All supplemental materials are available on OSF at \url{https://osf.io/uke76/?view_only=2ab48fcf974e47a2a9d1f7226e1b3c8b}, released under a CC BY 4.0 license. In particular, they include (1) the dataset we used for our qualitative exploration, (2) the \verb|combinations.js| file that we used to apply the constraints and generate the viable combinations, (3) a pdf  file providing more details on the implementation of our tool, showcasing usage examples, and including a link to the code repository, (4) a website with the 168 visualizations that are included in our design space, also accessible via:  \url{https://www.omvs-designspace.com/}, (5) csv files with the results of our qualitative evaluation, (6) the website developed for the crowdsource evaluation, (7) the anonymized experimental data, (8) the Jupiter notebooks used for the analysis, (9) the tables with the results, (10) the implementations of \EplusM and Facet.

\bibliographystyle{ACM-Reference-Format}
\bibliography{omv-experiment.bib}
\end{document}